\documentstyle[prb,aps,eqsecnum,multicol]{revtex}

\newcommand{\bleq}{\ifpreprintsty
                   \else
                   \end{multicols}\vspace*{-3.5ex}{\tiny
                  \noindent\begin{tabular}[t]{c|}
                  \parbox{0.493\hsize}{~} \\ \hline \end{tabular}}
                   \fi}
\newcommand{\eleq}{\ifpreprintsty
                 \else
                   {\tiny\hspace*{\fill}\begin{tabular}[t]{|c}\hline
                    \parbox{0.49\hsize}{~} \\
                    \end{tabular}}\vspace*{-2.5ex}\begin{multicols}{2}
                    \fi}
\newcommand{\bcols}{\ifpreprintsty\else\begin{multicols}{2}\fi}
\newcommand{\ecols}{\ifpreprintsty\else\end{multicols}\fi}

\begin{document}
\draft

\title{Effective action and collective modes in quasi-one-dimensional
spin-density-wave systems }

\author{K. Sengupta$^{(1)}$ and N. Dupuis$^{(2)}$  }
\address{(1) Department of Physics, University of Maryland, College Park, MD
20742-4111, USA \\
(2) Laboratoire de Physique des Solides, Universit\'e Paris-Sud, 91405
Orsay, France }
\date{January 24, 2000}
\maketitle

\begin{abstract}
We derive the effective action describing the long-wavelength
low-energy collective modes of quasi-one-dimensional spin-density-wave
(SDW) systems, starting from the Hubbard model within weak coupling
approximation. The effective action for the spin-wave mode corresponds
to an anisotropic non-linear sigma model together with a Berry phase
term. We compute the spin stiffness and the spin-wave velocity.  We
also obtain the effective action for the sliding mode (phason) taking
into account the density fluctuations from the outset and in presence
of a weak external electromagnetic field. This leads to coupled
equations for the phase of the SDW condensate and the charge density
fluctuations. We also calculate the conductivity and the
density-density correlation function.
\end{abstract}

\pacs{PACS Numbers:72.15.Nj, 75.30.Fv}

\bcols

\section{Introduction}

Materials with highly anisotropic crystal structure often exhibit
density-wave (DW) instabilities at low temperature.\cite{Gruner94}
While inorganic linear chain compounds usually develop a
charge-density-wave (CDW) instability, several group of organic
conductors present a SDW ground state. Well known examples include
transition metal bronzes such as K$_{0.3}$MoO$_3$ for CDW systems, and
the Bechgaard salts (TMTSF)$_2$X (X=ClO$_4$,PF$_6$) for SDW systems.

In quasi-one-dimensional (quasi-1D) systems, DW instabilities result
from nesting properties of the Fermi surface. Indeed, in this case the
particle-hole response function exhibits a logarithmic singularity
$\sim \ln(E_0/T)$, where $E_0$ is an ultra-violet cutoff of the order
of the bandwidth. In presence of repulsive electron-electron or
electron-phonon interactions, this leads to an instability of the
metallic phase at low temperature. CDW ground states resulting from
electron-phonon interaction were first discussed by Fr\"ohlich
\cite{Frohlich54} and Peierls,\cite{Peierls55} while the possibility
of a SDW ground state due to repulsive electron-electron interaction
was first postulated by Overhauser. \cite{Overhauser62}

In the SDW ground state, quasi-particle excitations exhibit a gap
$2\Delta_0$, where $\Delta_0$ is the SDW order parameter. The
low-energy behavior of the system is then dominated by low-lying
collective modes. In presence of an incommensurate SDW, two continuous
symmetries are spontaneously broken: the translational symmetry and
the rotational symmetry in spin space. This leads to the existence of
two gapless Goldstone modes: a sliding mode (phason) and a spin-wave
mode (magnon).  Contrary to the case of superconductors, collective
modes in DW systems directly couple to external fields, so that they
can easily be observed in various experiments. For instance, the
non-linear dc conductivity is a manifestation of the existence of a
phason mode which is pinned by impurities in real systems.

In CDW systems, collective modes were first studied by Lee, Rice and
Anderson.\cite{Lee74} From the computation of the Green functions,
they deduced the existence of a gapped amplitude mode and a gapless
phase mode (phason). The latter corresponds to a sliding of the CDW
and would lead to an infinite conductivity in a clean system ({\it
i.e.} with no impurity) as first proposed by
Fr\"ohlich. \cite{Frohlich54} Fukuyama then proposed an effective
phase Lagrangian ${\cal L}(\theta)$ determining the dynamics of the
phason,\cite{Fukuyama76} where the condensate phase $\theta$
determines the position of the DW with respect to the underlying
crystal lattice. The first attempt to derive ${\cal L}(\theta)$
rigorously is due to Brazovskii and Dzyaloshinskii. \cite{Brazovskii76}
The phase Lagrangian was latter used to study the interaction of the
CDW with impurities and the mechanism of pinning/depinning which is at
the origin of the non-linear dc conductivity observed in transport
experiments.  \cite{Fukuyama78,Lee79}

To a large extent, the analysis of collective modes in CDW systems can
be transposed to SDW systems.
\cite{Lee74,Psaltakis84,Maki90a,Maki90b,Brazovskii93,Poilblanc87,Zanchi,Maki87}
The amplitude mode has a gap $2\Delta_0$ and is therefore strongly
damped due to the coupling with quasi-particle excitations above the
mean-field gap. \cite{Psaltakis84} As in CDW systems, the phason
corresponds to a sliding of the DW and leads to an infinite Fr\"ohlich
conductivity in the absence of impurities. However, since the SDW
instability is driven by electron-electron interactions, the coupling
to the lattice plays no role and its mass is not renormalized by
phonons. The pinning by impurities is also weaker, since the SDW
couples to charge inhomogeneities only to second order. The spin-wave
mode, which is specific to SDW system, is also obtained from the poles
of the Green functions. \cite{Poilblanc87,Zanchi,Maki87}

The functional integral formalism has also proven useful to study
collective modes in DW systems.
\cite{Brazovskii76,Takano82,Krive85,Su86,Ishikawa88,Suzumura90,Girard93,Nagaosa96,Yak98}
The main advantage of this formalism is that it allows to derive the
effective Lagrangian of the Goldstone modes from first
principles. Both the phase Lagrangian  ${\cal L}(\theta)$ and the effective
Lagrangian for the spin-wave mode can be obtained within this formalism.

In this paper we derive the effective action describing the low-energy
collective modes of quasi-1D SDW systems, starting from the Hubbard
model with weak on-site interaction. This effective action describes the
behavior of the system at energies much smaller than the mean-field gap
$\Delta_0$ or, equivalently, at length scales much larger than the mean-field
coherence lengths $v_F/\Delta_0$ and $v_\perp/\Delta_0$ ($v_F$ and $v_\perp$
are the velocities along and across the conducting chains).
The improvement with respect to
previous works is twofold. First, we show that the effective action
for the spin-wave mode is given by an anisotropic non-linear sigma
model (NL$\sigma$M) together with a Berry phase term. Such a result is
known for the isotropic 2D Hubbard model,\cite{Wen88} but has not been
derived for weakly coupled chain systems. Instead, it is generally
assumed that the spin dynamics can be deduced from an effective
Heisenberg Hamiltonian. \cite{Gruner94} Second, we introduce the
long-wavelength charge-density field from the outset. We thus obtain
an effective Lagrangian ${\cal L}(\theta,\rho)$ which is a functional
of two independent fields: $\theta$, which is the phase of the SDW
condensate and the charge-density field $\rho$. This yields coupled
equations of motion for $\theta$ and $\rho$. The interaction between
these two quantities leads to a renormalization of the longitudinal
phason velocity.

The first step in the functional integral formalism is to introduce
auxiliary fields describing spin and charge fluctuations. The main
technical difficulty is then to recover the mean-field (or
Hartree-Fock) solution in a saddle-point approximation, while
maintaining rotational invariance in spin space which is a necessary
condition for obtaining the spin-wave mode.  To overcome this
difficulty, we introduce a space and time fluctuating
spin-quantization axis, following a method introduced by
Schulz\cite{Schulz90} and Weng {\it et al.}  \cite{Ting91} for the
isotropic 2D Hubbard model. Note that in quasi-1D systems, the
distinction between right and left moving electrons allows one to write
the Hamiltonian in a rotationally invariant form which is well suited
for the calculation of the spin-wave mode if one focuses only on the
$2k_F$ particle-hole (Peierls) channel. \cite{note1} However, the
concomitant consideration of the Landau channel (long-wavelength
charge fluctuations), which is at the heart of our approach, does
require the introduction of a fluctuating spin-quantization axis.

The organization of the paper is as follows. The effective action of
the system is derived in Sec.~\ref{secea}. We first introduce bosonic
fields describing charge and spin fluctuations, and the fluctuating
spin-quantization axis. We take special care to introduce the physical
charge density field $\rho$. The standard mean-field theory is
recovered in Sec.~\ref{mft}, while fluctuations are studied in
Sec.~\ref{ssecf}.  The latter are most conveniently computed by
performing a chiral rotation of the Fermion fields. The corresponding
Jacobian (the so-called chiral anomaly) is calculated in
Sec.~\ref{sssecca}.

The effective action governing the dynamics of the spin-wave mode is
shown to be a NL$\sigma$M together with a topological Berry phase term
in Sec.~\ref{secswm}. We explicitly calculate the spin stiffness, the
spin-wave velocity, and the coupling constant of the NL$\sigma$M.

The sliding mode is studied in Sec.~\ref{secpm}. We obtain coupled
equations of motion for the phase $\theta$ of the SDW condensate and
the charge fluctuations $\rho$. By integrating out one of these
fields, we obtain the effective action as a functional of either
$\theta$ or $\rho$. We also calculate the conductivity and the
density-density correlation function from the effective action.

We do not consider long-range Coulomb interaction, which would require
taking into account normal electrons that are thermally excited above
the gap. The latter are indeed expected to play a crucial role in the
screening of the interaction.\cite{Virosztek94} Note however that the
Coulomb interaction affects neither the spin-wave mode nor the
transverse phason mode sampled by optical spectroscopy.

We consider only the zero-temperature limit, and take $\hbar=k_B=1$
throughout the paper.

\section{Effective action}
\label{secea}

In the vicinity of the Fermi level, the electron dispersion is well
approximated as
\begin{equation}
\epsilon(k_x,k_y) = v_F(|k_x|-k_F)-2t_b \cos (k_yb),
\label{disp}
\end{equation}
where $k_x$ and $k_y$ are the electron momenta along and across the
conducting chains, $t_b$ is the transfer integral in the transverse
direction, and $b$ the inter-chain distance. In Eq.~(\ref{disp}), the
longitudinal dispersion is linearized around the 1D Fermi points $\pm
k_F$ and $v_F=2at_a\sin(k_Fa)$ is the corresponding Fermi velocity,
with $t_a\gg t_b$ being the transfer integral and $a$ the lattice
spacing along the chains. The linearized dispersion (\ref{disp})
satisfies the property $\epsilon({\bf k})=-\epsilon({\bf k}+{\bf Q})$,
which corresponds to a perfect nesting of the Fermi surface at wave
vector ${\bf Q}=(2k_F,\pi/b)$.  Actually this property is an artifact
of the linearization and does not hold for the original tight-binding
dispersion unless the system is half-filled ($k_F=\pi/2a$). Deviations
from perfect nesting can be taken into account by adding higher
harmonics to the transverse dispersion $-2t_b\cos(k_yb)$. For
simplicity, we shall not consider such terms and restrict ourselves to
the perfect nesting case. We also assume that the SDW is
incommensurate with the crystal lattice, so that there is no pinning
by the lattice.

Following the standard procedure, we introduce right and left
Fermionic fields $\psi_{+\sigma}$ and $\psi_{-\sigma}$ ($\sigma$ is
the spin index). In terms of the quartet of Fermion fields
$\psi^\dagger=(\psi^\dagger_{+\uparrow},\psi^\dagger_{-\uparrow},
\psi^\dagger_{+\downarrow},\psi^\dagger_{-\downarrow})$, one can write
the Hamiltonian of the system as $H=H_0+H_I$, where
\begin{eqnarray}
H_0 &=& \sum_{n} \int \,dx \, \Big[
\psi^{\dagger}_n(x)
 v_F(-i\partial_x \tau_3 - k_F )  \psi_{n}(x) \nonumber\\ &&
- t_b \sum_{\delta = \pm 1}
\psi^{\dagger}_{n} (x) \psi_{n+\delta}
(x) \Big] ,\nonumber\\
H_I &=& U \sum_{n} \int \,dx \,\psi^{\dagger}_{n
\uparrow} (x) \,\psi^{\dagger}_{n \downarrow}(x)
\psi_{n \downarrow}(x) \psi_{n \uparrow}(x),
\label{hub1}
\end{eqnarray}
$U$ is the on-site Coulomb interaction strength and $n$ the chain
index. Here and in the following, $\tau_\mu$ ($\mu=1,2,3$) and
$\sigma_\nu$ ($\nu=x,y,z$) are $2\times2$ Pauli matrices acting on
left/right and spin indices of the Fermionic fields respectively. The
product $\tau_\mu\sigma_\nu$ is to be understood as direct product of
the matrices $\tau_\mu$ and $\sigma_\nu$, and any single matrix
$\tau_\mu$ or $\sigma_\nu$ as direct product of that matrix with the
unit matrix.

Introducing the charge and spin density fields
\begin{eqnarray}
\rho_{c} &=& \psi^{\dagger}\psi , \,\,\,\,\,\,
\rho_{s} = \psi^{\dagger}\sigma_z\psi ,  \nonumber\\
\rho_{c+} &=& \psi^{\dagger}\tau_-\psi, \,\,\,\,\,\,
\rho_{s+} = \psi^{\dagger}\tau_- \sigma_z \psi ,
\end{eqnarray}
where $\tau_{\pm}=(\tau_1\pm i\tau_2)/2$, we rewrite the interaction
Hamiltonian  as
\begin{eqnarray}
H_I &=& \frac{U}{4}\sum_{n} \int \,dx \,
\big[ (\rho_{c n}(\tau,x))^2 - (\rho_{s n}(\tau,x))^2 \nonumber\\
&& + 2 \left( \rho_{c +\,n}^{*}(\tau,x)\rho_{c +
\,n}(\tau,x) \right. \nonumber\\
&& \left. -\rho_{s +\,n}^{*}(\tau,x) \rho_{s +\,n}
(\tau,x)\right) \big ].
\label{hub2}
\end{eqnarray}

In the Matsubara formalism, the partition function of the system can
be written as a functional integral over anticommuting Grassmann
variables, $Z=\int D\psi^{\dagger} D\psi e^{-S[\psi^{\dagger},\psi]}$,
with the action
\begin{equation}
S = \int_{0}^{\beta} d\tau \biggl[ \sum_{n} \int dx  \,
\psi^{\dagger}_n(\tau,x) \partial_{\tau}
\psi_{n}(\tau,x) + H [\psi^\dagger,\psi] \biggr] ,
\label{ac1}
\end{equation}
where $\beta=1/T$ is the inverse temperature. The limit $T\to 0$ is to
be taken at the end of the calculations.

At this point, it is customary to introduce auxiliary fields for the
spin and charge fluctuations via a Hubbard-Stratonovitch (HS)
transformation.  However, as noted earlier \cite{Schulz90,Ting91} in
the context of systems described by the isotropic 2D Hubbard model,
such a procedure immediately leads to loss of spin rotational
invariance. The reason for this is that in writing down the
Hamiltonian, we have made a particular choice ($\hat z$) for the
spin-quantization axis of the electrons. Other decompositions of $H_I$ in terms
of charge and spin fluctuations are possible. They are all equivalent as far as
the partition function is calculated exactly. The reason for choosing the
decomposition (\ref{hub2}) is that it allows to recover the Hartree-Fock
solution at the saddle point level within the functional integral formalism
(see Sec.~\ref{mft}).

In order to maintain spin
rotational invariance, one should consider the spin-quantization axis
to be a priori arbitrary and integrate over all possible directions in
the partition function. This is done in practice by introducing a new
field $\phi_n(\tau,x)$ which is related to the old Fermionic field
$\psi_n(\tau,x)$ through an unitary SU(2)/U(1) rotation matrix
$R_n(\tau,x)$, {\it i.e.} $ \psi = R \phi$. The rotation matrix $R$
satisfies $ R\sigma_zR^\dagger = \sigma\cdot {\bf n}$, where ${\bf n}$
is an unit vector field which gives the direction of spin-quantization
axis at point $(x,n)$ and time $\tau$.  The new field $\phi$ has its
spin-quantization axis along the local ${\bf n}$ vector. The
interaction term in the action, $S_I$ [Eqs.~(\ref{hub2}) and
(\ref{ac1})], is invariant under this transformation while the
unperturbed part, $S_0$, becomes
\begin{eqnarray}
S_0 &=& \sum_{n} \int_{0}^{\beta} d\tau \,dx \,{\Big \{}
\phi^{\dagger}_n(\tau,x)[ \partial_{\tau} - C_0 \nonumber\\ &&
+v_F(-i\partial_x\tau_3 -k_F ) - v_F \tau_3 C_x ] \phi_n(\tau,x)
\nonumber\\ && - t_b \sum_{\delta = \pm 1} \phi^{\dagger}_{n}(\tau,x)
e^{-i\int^{(n+\delta)b}_{nb} C_y dy } \phi_{n+\delta} (\tau,x){\Big \}}.
\label{ac2}
\end{eqnarray}
We have introduced the fields $C_{\mu}$ given by
\begin{eqnarray}
C_0 &=& -R^{\dagger} \partial_{\tau} R ,\nonumber\\ C_x &=& i
R^{\dagger} \partial_x R ,\nonumber \\ e^{-i\int^{(n+\delta)b}_{nb} C_y
dy } &=& R^{\dagger}_{n} R_{n+\delta}.
\label{gf1}
\end{eqnarray}
The fields $C_{\mu}$ are SU(2) gauge fields and the U(1) gauge freedom
here corresponds to an arbitrary rotation about the $z$ axis, which
does not change the state of the system. The partition function now
contains an additional integral over all $C_{\mu}$ or equivalently
${\bf n}$ field configurations. The $C_{\mu}$ fields thus contain
information about the spin excitations of the system. In writing
Eqs.~(\ref{ac2}) and (\ref{gf1}), we have considered only
long-wavelength fluctuations of the spin-quantization axis, so that
the matrix $R$ acts like the unit matrix with respect to the
left/right indices of the Fermionic fields.

The $2k_F$ charge fluctuations play no role in the SDW phase, so that
we ignore the term $\propto \rho_{c +\,n}^{*}\rho_{c +\,n}$ in the
interaction Hamiltonian $H_I$. We thus write the interaction term in
the action as
\begin{equation}
S_I = \frac{U}{4}\sum_{n} \int_{0}^{\beta} d\tau\, dx \,
\left[ (\rho_{c n}(\tau,x))^2
 - 2 \rho_{s +}^{*}(\tau,x)\rho_{s +} (\tau,x)\right].
\end{equation}
Note that we have kept explicitly the long-wavelength charge
fluctuations since they couple to the phason mode.\cite{Gruner94} In
principle, one should also retain the long-wavelength spin
fluctuations which couple to the spin-wave mode. This coupling
renormalizes the spin-wave velocity by the usual Stoner
factor.\cite{Gruner94} In practice, it seems difficult within our
formalism to treat the long-wavelength spin fluctuations in a way that
preserves spin rotation invariance. For this reason, we shall ignore
them in the following.

We introduce two HS fields, $\rho^{\rm HS}$ (real) and $\Delta$
(complex), corresponding to charge and spin density fluctuations,
respectively. The action of the system can then be written as $S = S_0
+ S_I$, where $S_0$ is given by Eq.~(\ref{ac2}), and $S_I$ by
\begin{eqnarray}
S_I &=& \sum_{n} \int_{0}^{\beta} d\tau\, dx \,{\Big\{}
 -i \rho_{c\,n}(\tau,x) \rho_{n}^{\rm HS}(\tau,x) \nonumber\\
&& - \left[\Delta_n^{*}(\tau,x) \rho_{s+\,n}(\tau,x)
+ {\rm c.c.} \right] \nonumber\\ &&
+ \frac{1}{U} \Big[(\rho_{n}^{\rm HS}(\tau,x))^2
+ 2 |\Delta_n(\tau,x)|^2 \Big] {\Big \}}.
\label{ac3}
\end{eqnarray}
Note that the HS field $\rho^{\rm HS}$ introduced here is not the
physical charge-density field, but its conjugate.  This can be easily
checked by varying the action $S$ with respect to $\rho^{\rm
HS}$. Following Palo ${\it et\,al.}$, \cite{Palo99} we now introduce
the physical charge density field $\rho$ by decoupling the quadratic
term in $\rho_{n}^{\rm HS}(\tau,x)$ by means of a HS
transformation. This leads to the action
\begin{eqnarray}
S_I &=& \sum_{n} \int_{0}^{\beta} d\tau\, dx
\Big[ -i\rho_{c\,n}(\tau,x) \rho_{n}^{\rm HS}(\tau,x)\nonumber\\
&& - \left[\Delta_n^{*}(\tau,x) \rho_{s+\,n} (\tau,x)
+{\rm c.c.} \right] + \frac{U}{4}(\rho_{n}(\tau,x))^2 \nonumber\\
&& + \frac{2}{U} |\Delta_n(\tau,x)|^2 + i \rho_{n}^{\rm
HS}(\tau,x)\rho_{n}(\tau,x) \Big] .
\label{ac4}
\end{eqnarray}
It can be seen, by varying the action with respect to the HS field
$\rho^{\rm HS}$, that $\rho$ is indeed the physical charge density
field.

Finally, we introduce a weak external electromagnetic field $A_{\mu}$
in the action in a gauge-invariant manner. This does not change $S_I$
while $S_0$ becomes
\begin{eqnarray}
S_0 &=& \sum_{n} \int_{0}^{\beta} d\tau\, dx \,{\Big \{}
\phi^{\dagger}_n(\tau,x)\Big[
\partial_{\tau}-e A_0 -C_0\nonumber\\
&&   +v_F\left((-i\partial_x -e A_x -C_x )\tau_3
- k_F \right)   \Big] \phi_n(\tau,x) \nonumber\\ &&
- t_b \sum_{\delta = \pm 1}\phi^{\dagger}_{n}(\tau,x)
e^{-i\int^{(n+\delta)b}_{nb} (C_y + eA_y )dy }
\phi_{n+\delta}(\tau,x){\Big \}}. \nonumber \\ &&
\label{ac5}
\end{eqnarray}
Note that, as an artifact of the linearization of the dispersion
relation along the chain, $S_0$ does not contain the diamagnetic term
$\propto A_x^2$. However, as we shall see, this term is important for
recovering a gauge-invariant effective action. We shall recover this
term starting from the original tight-binding Hamiltonian in
Sec.~\ref{ssecf}.

\subsection{Mean-field theory}
\label{mft}

The standard mean-field theory is recovered from a saddle-point
approximation with ${\bf n} = {\hat z}$, $C_{\mu}= A_{\mu} = 0$, $\rho
= \rho_0$, $\rho^{\rm HS} = \rho_0^{\rm HS}$ and $\Delta = \Delta_0
\exp(i{\bf Q}\cdot{\bf r})$, where ${\bf Q}$ is the nesting
wave-vector and ${\bf r}=(x,nb)$. We take $\Delta_0$ to be real,
without any loss of generality. Varying the action with respect to the
auxiliary fields $\rho$, $\rho^{\rm HS}$ and $\Delta$, we obtain the
mean-field equations
\begin{eqnarray}
\rho_0 &=& -\frac{2i}{U} \rho_0^{\rm HS}
= \langle \phi^{\dagger}_n(\tau,x) \phi_n(\tau,x) \rangle_{\rm MF} ,\nonumber\\
\Delta_0 &=& \frac{U}{2} e^{-i {\bf Q} \cdot {\bf r}}
\langle \phi^{\dagger}_n(\tau,x) \tau_-  \sigma_z
\phi_n(\tau,x) \rangle_{\rm MF}.
\label{mfe}
\end{eqnarray}
It is clear from the above equations that the mean field $\rho_0$ is
the density of particles in the system and has the same value in both
the metallic and the SDW ground state. As a result, $-i\rho_0^{\rm
HS}$ can be absorbed as a trivial shift in the chemical potential of
the system [see Eq.~(\ref{ac4})]. The equation for $\Delta_0$ is the
usual mean-field equation for the SDW order parameter.

The mean-field propagator can now be obtained from the mean-field
Fermionic action
\bleq
\begin{eqnarray}
S_{\rm MF} &=& - \frac{1}{\beta}\sum_{\omega_n}\sum_{{\bf k},\sigma}
\left(\phi_{+ \sigma}^{\dagger}(i \omega_n,{\bf k}+{\bf Q}),
\phi_{- \sigma}^{\dagger}(i \omega_n,{\bf k})\right) {\mathcal G}_
{\sigma}^{-1}{\phi_{+ \sigma}(i \omega_n,{\bf k}+{\bf Q}) \choose
\phi_{- \sigma}(i \omega_n,{\bf k})} ,
\label{mfa}
\end{eqnarray}
\eleq
where we have dropped additive contributions to the free energy due to
the mean-fields $\rho_0$, $\rho^{\rm HS}_0$ and $\Delta_0$.  Here
${\mathcal G}_{\sigma}^{-1}$ is the inverse propagator given by
\begin{eqnarray}
{\mathcal G}_{\sigma}^{-1} &=&
\left( \begin{array}{cc}
i \omega_n -\epsilon_+({\bf k}+{\bf Q}) & {\rm sgn}(\sigma) \Delta_0 \\
{\rm sgn}(\sigma) \Delta_0 & i \omega_n -\epsilon_-({\bf k})
\end{array} \right) ,
\end{eqnarray}
where ${\rm sgn}(\sigma) = +$($-$) for $\sigma =
\uparrow$($\downarrow$) and $\epsilon_{\pm}({\bf k}) = v_F({\pm}k_x
-k_F) -2t_b \cos(k_y b)$ is the dispersion relation for the right and
left Fermions.

The mean-field propagator is obtained by inverting ${\mathcal
G}_{\sigma}^{-1}$. It is given by
\begin{eqnarray}
{\mathcal G}_{\sigma} &=& \left( \begin{array}{cc}
G_{+\sigma}(i\omega_n,{\bf k}+{\bf Q})& F_{+\sigma}(i\omega_n,{\bf k}+{\bf
Q}) \\ F_{-\sigma}(i\omega_n,{\bf k})& G_{-\sigma}(i\omega_n,{\bf k})
\end{array} \right),
\label{mfgf}
\end{eqnarray}
where
\begin{eqnarray}
G_{\pm \sigma} (i \omega_n,{\bf k})
&=& - \langle \phi_{\pm \sigma}
(i \omega_n,{\bf k})\phi_{\pm \sigma}^{\dagger}
(i \omega_n,{\bf k}) \rangle \nonumber\\
&=& -\frac{i\omega_n + \epsilon_{\pm}({\bf
k})}{\omega_n^2 + \epsilon_{\pm}^2 +\Delta_0^2}, \nonumber\\
F_{\pm \sigma}(i \omega_n,{\bf k}) &=& - \langle \phi_{\pm \sigma}
(i \omega_n,{\bf k})\phi_{\mp \sigma}^{\dagger}
(i \omega_n,{\bf k}\mp {\bf Q} )\rangle \nonumber\\
&=& \frac{{\rm sgn}(\sigma) \Delta_0}
{\omega_n^2 + \epsilon_\pm^2 +\Delta_0^2},
\label{gfc}
\end{eqnarray}
and we have used the relation $\epsilon_+({\bf k}+{\bf Q}) =
-\epsilon_-({\bf k}) $ in obtaining the above result.

\subsection{Fluctuations}
\label{ssecf}

We do not consider amplitude fluctuations, since they are gapped and
decouple from the sliding and spin-wave modes in the long-wavelength
limit.\cite{Gruner94} We therefore write the auxiliary fields as
\begin{eqnarray}
\rho^{\rm HS} &=&
\rho_0^{\rm HS}+\delta \rho^{\rm HS}, \nonumber\\
\rho &=& \rho_0 +\delta \rho ,\nonumber\\
\Delta &=& \Delta_0 e^{i({\bf Q}\cdot {\bf r} + \theta)},
\end{eqnarray}
where $\delta \rho$, $\delta \rho^{\rm HS}$, and $\theta$ represent
small fluctuations of the fields about their mean-field values. The
action can then be written as $S=S'_0 + S'_I$, with
\begin{eqnarray}
S'_0 &=& \sum_{n} \int_{0}^{\beta} d\tau\, dx \,{\Big \{}
\phi^{\dagger}_n(\tau,x)\Big[
\partial_{\tau}-e A_0 - C_0 -i \delta \rho^{\rm HS} \nonumber\\
&& +v_F\left((-i\partial_x -e A_x -C_x )\tau_3
- k_F \right) \nonumber\\ && -\Delta_0
\Big( e^{-i({\bf Q}\cdot {\bf r} + \theta)}
\tau_-  \sigma_z + {\rm c.c.} \Big) \Big] \phi_n(\tau,x) \nonumber\\
&& - t_b \sum_{\delta = \pm 1 }\phi^{\dagger}_{n}(\tau,x)
 e^{-i\int^{(n+\delta)b}_{nb} ( C_y + eA_y )dy }
\phi_{n+\delta}(\tau,x){\Big \}} ,\nonumber\\
S'_I &=& \sum_{n} \int_{0}^{\beta} d\tau\, dx \,
 \Big[\frac{U}{4}(\delta \rho_{n}(\tau,x))^2
+i \delta \rho_{n}^{\rm HS}(\tau,x) \nonumber\\
&& \times \Big(\rho_0
+ \delta \rho_{n}(\tau,x) \Big) \Big].
\label{ac6}
\end{eqnarray}

The next step to obtain the effective action of the auxiliary fields
$\delta\rho$, $\delta\rho^{\rm HS}$ and $\theta$ is to integrate out
the Fermion fields. This is most conveniently done by first
introducing a new field $\phi^{'}$ related to the field $\phi$ through
an unitary chiral transformation:
\begin{eqnarray}
\phi &=& U_{\rm chiral} \phi^{'} = e^{i\tau_3 \theta/2} \phi^{'}.
\label{Uchiral}
\end{eqnarray}
This transformation leaves the interaction part of the action $S'_I$
invariant while $S'_0$ is given by
\begin{eqnarray}
S'_0 &=& \sum_{n} \int_{0}^{\beta} d\tau\, dx\,{\Big \{}
\phi^{'\,\dagger}_n(\tau,x)\Big[
\partial_{\tau} - A_0^{\rm tot} \nonumber\\
&& +v_F\left(-i\partial_x \tau_3 - k_F
- A_x^{\rm tot} \right) \nonumber\\
&& -\Delta_0 \Big( e^{-i{\bf Q}\cdot {\bf r}}
 \tau_-  \sigma_z + {\rm c.c.} \Big) \Big]
\phi_n^{'}(\tau,x) \nonumber\\
&& - t_b \sum_{\delta = \pm 1}\phi^{'\,\dagger}_{n}(\tau,x)
e^{-i\int^{(n+\delta)b}_{nb}
A_y^{\rm tot} dy } \phi^{'}_{n+\delta}(\tau,x) {\Big \}},\nonumber\\
\label{ac7}
\end{eqnarray}
where the gauge field $A_{\mu}^{\rm tot}$ is given by
\begin{eqnarray}
A_{0}^{\rm tot}&=& e A_0 -\frac{v_F}{2} \partial_x \theta
+ i \delta\rho^{\rm HS} +  C_0 ,\nonumber\\
A_{x}^{\rm tot}&=& (e A_x -\frac{i}{2 v_F} \partial_{\tau}
\theta  + C_x ) \tau_3 ,\nonumber\\
A_{y}^{\rm tot}&=& e A_y  - \frac{1}{2} \partial_y \theta
\tau_3 + C_y.
\label{tgf1}
\end{eqnarray}
The chiral transformation therefore eliminates the phase $\theta$ of
the order parameter $\Delta$. The action $S'_0$ [Eq.~(\ref{ac7})]
therefore acquires a simple form. The tradeoff, however, is that the
Fermions are now subjected to an effective potential $A_{\mu}^{\rm
tot}$ which contains the derivatives of the phase field
$\theta$. Notice that in contrast to the superconducting case, the
gradients of the phase couple to the external electromagnetic field in
a different manner. This is a consequence of the different broken
symmetries in the two cases.  Since we assume that the fluctuations of
the order parameter (we have taken $\theta_{\rm MF} = 0$) are small,
the gradients of $\theta$ must be small. As a result, one can carry
out a perturbative expansion of the action in $\partial_{\mu} \theta$
or equivalently in $A_{\mu}^{\rm tot}$. The chiral transformation also
produces a non-trivial Jacobian $J$, which yields the additional
contribution $S_J =-\ln J$ to the action. We shall come back to the
origin of this non trivial Jacobian and the method of its calculation
in more detail in Sec.~\ref{sssecca}.

We are now in a position to write the action in terms of the
mean-field action $S_{\rm MF}$ [Eq.~(\ref{mfa})] and the part
involving fluctuations. The action then reads $S = S'_0 + S'_I+S_J$,
where
\begin{eqnarray}
S'_0 &=& S_{\rm MF} \nonumber \\ &&
- \sum_{n} \int_{0}^{\beta} d\tau\, dx\,{\Big \{}
\phi^{'\,\dagger}_n(\tau,x)\left[
 A_0^{\rm tot} + v_F A_x^{\rm tot} \right]   \phi_n^{'}(\tau,x)
 \nonumber \\ &&
- t_b \sum_{\delta = \pm 1} \phi^{'\,\dagger}_{n}(\tau,x)
\Big(e^{-i\int^{(n+\delta)b}_{nb}
A_y^{\rm tot} dy } - 1\Big)  \phi^{'}_{n+\delta}(\tau,x){\Big \}}.
\nonumber \\
\label{ac8}
\end{eqnarray}

Using the fact that $A_y^{\rm tot}$ is a slowly varying weak field, we
expand the factor $\exp(-i\int^{(n+\delta)b}_{nb} A_y^{\rm tot} dy
)-1$ in $S'_0$ in powers of $A_y^{\rm tot}$. The terms in the
expansion are written in a symmetric way with respect to $n$ and
$n+\delta$. Here, we retain terms up to the quadratic order in
$A_y^{\rm tot}$. The corresponding contribution to the action reads
$S_y=S_y^{\rm linear}+S_y^{\rm dia}$, where
\begin{eqnarray}
S_{y}^{\rm linear} &=& \frac{it_b b}{2}
\sum_{n,\delta=\pm} \int_{0}^{\beta} d\tau\, dx \,
\phi^{'\,\dagger}_{n}(\tau,x)
\Bigl( A_{y\,n}^{\rm tot}(\tau,x) \nonumber\\ &&
+ A_{y\,n+\delta}^{\rm tot}(\tau,x) \Bigr) \delta
\phi^{'}_{n+\delta}(\tau,x) , \nonumber\\
S_{y}^{\rm dia} &=& \frac{t_b b^2}{8} \sum_{n,\delta=\pm}
\int_{0}^{\beta} d\tau \, dx \,
\phi^{'\,\dagger}_{n}(\tau,x) \Big( A_{y \,n}^{\rm
tot}(\tau,x) \nonumber\\
&& + A_{y \,n +\delta }^{\rm tot}(\tau,x) \Big)^2
\phi^{'}_{n+\delta}(\tau,x) .
\label{ydia}
\end{eqnarray}
The linear term $S_{y}^{\rm linear}$ corresponds to the coupling of
$A_y^{\rm tot}$ to the paramagnetic current in the transverse
direction, while $S_y^{\rm dia}$, which is quadratic in $A_y^{\rm
tot}$, is the diamagnetic contribution. If we had started with the
original tight-binding Hamiltonian in the $x$ direction, we would have
come up, apart from the linear paramagnetic term, with a similar
diamagnetic term in the $x$ direction given by
\begin{eqnarray}
S_{x}^{\rm dia}&=& \frac{t_a a^2}{2}\sum_{n}
\int_{0}^{\beta} d\tau \,\frac{dk_x dp_x}{(2\pi)^2}
\phi^{'\,\dagger}_{n}(k_x,\tau)
\nonumber \\ && \times |eA_{x\,n}(p_x,\tau)+C_{x\,n}(p_x,\tau)|^2
\nonumber \\ && \times
[ \cos(k_x a) + \cos((k_x-p_x)a)]\phi^{'}_{n}(k_x,\tau),
\label{xdia}
\end{eqnarray}
where $t_a$ is the hopping parameter along the chain. $S_x^{\rm dia}$,
which is not obtained if one uses a linear dispersion law from the
very beginning, has to be included in the action in order to maintain
gauge invariance.

It is convenient to introduce the charge and spin currents for the
Fermions:
\begin{eqnarray}
j_0^{\alpha} &=& \phi^{'\,\dagger} \sigma^{\alpha}\phi^{'}  ,\nonumber\\
j_x^{\alpha} &=& \phi^{'\,\dagger} v_F \tau_3 \sigma^{\alpha}\phi^{'}
,\nonumber\\
j_y^{\alpha} &=& -\frac{it_b b}{2}\sum_{\delta = \pm 1}
\delta (\phi^{'\,\dagger}_{n-\delta} \sigma^{\alpha} \phi^{'}_n
+\phi^{'\,\dagger}_{n} \sigma^{\alpha} \phi^{'}_{n+\delta}) ,\nonumber\\
g &=& -\frac{it_b b}{2}\sum_{\delta =\pm 1}
\delta (\phi^{'\,\dagger}_{n-\delta}\tau_3 \phi^{'}_n
+\phi^{'\,\dagger}_{n} \tau_3  \phi^{'}_{n+\delta}),
\label{scc}
\end{eqnarray}
where the index $\alpha$ runs over $0,x,y,z$, $\sigma^0$ is the unit
matrix, $j_{\mu}^0$ is the $\mu$ th component of the charge current,
and $j_{\mu}^{\alpha}$ for $\alpha \ne 0$ give different components of
the spin current. The current $g$ is almost same as the charge current
in the transverse direction, except that it is chiral, {\it i.e.} it
has opposite sign for the left and right moving Fermions. This current
is introduced for notational convenience, as we shall see later.

Furthermore, since the $C_{\mu}$ fields are SU(2) gauge fields, it is
possible to write them in terms of the $\sigma$ matrices, namely,
$C_{\mu} = A_{\mu}^{\nu}\sigma_{\nu}$, where the index $\mu$ runs over
$0$,$x$,$y$ and the index $\nu$ over $x,y,z$. As a result, the fields
$A_{\mu}^{\rm tot}$ can be expressed as
\begin{eqnarray}
A_{0}^{\rm tot}&=& A_0^0 + A_0^{\nu} \sigma_{\nu}, \nonumber\\
A_{x}^{\rm tot}&=& (A_x^0 + A_x^{\nu}\sigma_{\nu}) \tau_3 , \nonumber\\
A_{y}^{\rm tot}&=& A_y^0 + A_y^{\nu} \sigma_{\nu}
-\frac{1}{2} \partial_y \theta \tau_3 ,
\label{tgf2}
\end{eqnarray}
where the expressions for $A_{\mu}^0$ can be deduced from
Eqs.~(\ref{tgf1}) and (\ref{tgf2}) to be
\begin{eqnarray}
A_{0}^{0}&=& e A_0 -\frac{v_F}{2} \partial_x \theta
+ i \delta\rho^{\rm HS},\nonumber\\
A_{x}^{0}&=& e A_x -\frac{i}{2 v_F} \partial_{\tau}
\theta , \nonumber\\
A_{y}^{0}&=& e A_y .
\label{tgf3}
\end{eqnarray}
The action [Eq.~(\ref{ac8})] can be conveniently expressed in terms of
the charge and spin currents as $S = S_{\rm MF} + S'' + S^{\rm dia} +
S'_I + S_J $, where
\begin{eqnarray}
S'' &=& -\sum_{n} \int_{0}^{\beta} d\tau\, dx\,
\left(  \sum_{\mu}^{0,x,y,}\sum_{\alpha}^{0,x,y,z}
A_{\mu}^{\alpha} j_{\mu}^{\alpha}
-\frac{1}{2} g \partial_y \theta \right), \nonumber\\
\label{eac0}
\end{eqnarray}
and $S^{\rm dia}= S_x^{\rm dia} + S_y^{\rm dia}$ is given by
Eqs.~(\ref{ydia}) and (\ref{xdia}).

Integrating out the Fermions, we obtain to quadratic order in the
fields $A_\mu^{\rm tot}$ the effective action
\begin{eqnarray}
S_{\rm eff} &=& S'_I + S_J + \bigg \langle  S''
+S^{\rm dia}  -\frac{S''^2}{2} \bigg \rangle_{\rm MF},
\label{eac1}
\end{eqnarray}
where $\langle \cdots \rangle_{\rm MF}$ means that the average is
taken with respect to the mean-field action $S_{\rm MF}$.

The evaluation of $\langle S'' \rangle_{\rm MF}$ is trivial. Only the
term involving $j_0^0$ contributes. We thus obtain
\begin{eqnarray}
{\big \langle} S''{\big \rangle}_{\rm MF}&=& -\rho_0
\int_{0}^{\beta} d\tau d^2 r  \left( i \delta \rho^{\rm HS}
+ eA_0 -\frac{v_F}{2} \partial_x \theta \right) . \nonumber\\
\label{lt1}
\end{eqnarray}
Here, we have taken the continuum limit at the end of the calculation
and replaced the sum over the chains by an integral in the $y$
direction. It can be easily seen from Eqs.~(\ref{ac6}) and (\ref{lt1})
that the first term in ${\big \langle} S''{\big \rangle}_{\rm MF}$
cancels the term which is linear in $\delta \rho^{\rm HS}$ in the
expression of $S'_I$. Thus, we finally obtain
\begin{eqnarray}
\langle S''\rangle_{\rm MF}
&=& -\rho_0  \int_{0}^{\beta} d\tau d^2 r
\left(eA_0 -\frac{v_F}{2} \partial_x \theta \right),
\label{lt2}
\end{eqnarray}
and
\begin{equation}
S'_I = \int_{0}^{\beta} d\tau \,d^2 r \,\Big[
 \frac{U}{4}(\delta\rho)^2 +i \delta\rho^{\rm HS} \delta\rho \Big].
\label{it}
\end{equation}

It may seem unphysical at first sight that the scalar potential in the
effective action couples to the constant mean-field density $\rho_0$
and not to the full density $\rho = \rho_0 + \delta\rho$, as one would
intuitively expect. However, one should bear in mind that one still
has to integrate over $\delta\rho^{\rm HS}$. As shown by Palo ${\it
et\,al.}$ \cite{Palo99} in the context of superconducting systems, by
redefining the field $\delta \rho^{\rm HS} \rightarrow \delta
\rho^{\rm HS}-(eA_0 -\frac{v_F}{2} \partial_x \theta )$, we
immediately get the coefficient of the scalar potential to be the full
density $\rho$ and not $\rho_0$.

In the next three sections, we evaluate the diamagnetic term $\langle
S^{\rm dia}\rangle_{\rm MF}$, the contribution $S_J$ arising from the
chiral anomaly, and $\langle S''^2 \rangle_{\rm MF}$.

\subsubsection{Diamagnetic contribution}

In this section, we calculate the contribution of the diamagnetic term
to the effective action. From Eq.~(\ref{ydia}), one can easily obtain
in Fourier space
\bleq
\begin{eqnarray}
\langle S_y^{\rm dia}\rangle_{\rm MF} &=& \frac{t_b b^2}{2}
\frac{1}{\beta} \sum_{p_n} \int \frac{d^2p}{(2 \pi)^2}\,
\left(e^2 |A_y(ip_n,{\bf p})|^2 + \sum_{\nu}^{x,y,z} |A_y^{\nu}(ip_n,{\bf
p})|^2 +\frac{p_y^2}{4} |\theta(ip_n,{\bf p}) |^2 \right)
\nonumber \\ && \times \frac{1}{\beta}
\sum_{\alpha=\pm,\sigma,\omega_n} \int \frac{d^2k}{(2 \pi)^2}
[\cos (k_y b) + \cos((k_y-p_y)b)] G_{\alpha\sigma}(i\omega_n,{\bf k}),
\label{ydiar1}
\end{eqnarray}
where we use $A_y^*(ip_n,{\bf p})=A_y(-ip_n,-{\bf p})$... for real
fields and $G_{\pm\sigma}$ is the mean-field propagator given by
(\ref{gfc}). In the limit $t_b\gg \Delta_0$,\cite{note2} we can ignore
the effect of the gap and replace $G_{\pm\sigma}$ by its value in the
metallic phase. To first order in $p_y^2$, we then have
\begin{eqnarray}
 \frac{1}{\beta}
\sum_{\alpha=\pm,\sigma,\omega_n} \int \frac{d^2k}{(2 \pi)^2}
[\cos (k_y b) + \cos((k_y-p_y)b)] G_{\alpha\sigma}(i\omega_n,{\bf k})
&=& 4 N(0) \left(1-\frac{p_y^2 b^2}{4} \right)
\int \frac{dk_y}{2\pi} \,\cos(k_y b) \nonumber \\ && \times
\int^{2t_b}_{-2t_b} d\epsilon \,
\Theta(-\epsilon + 2t_b \cos (k_y b)),\nonumber\\
&=& \frac{2 N(0)v_{\perp}^2}{t_b b^2} \left(1-\frac{p_y^2 b^2}{4} \right),
\label{ydiar2}
\end{eqnarray}
where $v_{\perp}=\sqrt{2t_b^2b^2}$ is the velocity of the Fermions in
the transverse direction, $N(0)=1/\pi v_Fb$ is the density of states
per spin, and $\Theta$ is the step function. Note that only the states
near the Fermi level ($|\epsilon|\leq 2t_b$) contribute to the
diamagnetic term $\langle S_y^{\rm dia} \rangle_{\rm MF}$, so that the
use of a linearized dispersion law is justified. Substituting
Eq.~(\ref{ydiar2}) in Eq.~(\ref{ydiar1}), we finally get
\begin{equation}
\langle S_y^{\rm dia}\rangle_{\rm MF} = N(0)v_{\perp}^2
\frac{1}{\beta} \sum_{p_n} \int \frac{d^2p}{(2 \pi)^2}
\left(1- \frac{p_y^2 b^2}{4} \right)
\left(e^2 |A_y(ip_n,{\bf p})|^2 + \sum_\nu^{x,y,z}|A_y^{\nu}(ip_n,{\bf
p})|^2+ \frac{p_y^2}{4} |\theta(ip_n,{\bf p})|^2 \right).
\label{ydiaf}
\end{equation}

The calculation of the diamagnetic term in the $x$ direction should be
done using the original tight-binding dispersion law, since it
involves electronic states deep in the Fermi sea which forbids the use
of a linearized dispersion law. From Eq.~(\ref{xdia}), we deduce
\begin{eqnarray}
\langle S_x^{\rm dia}\rangle_{\rm MF} &=& \frac{t_a a^2}{2}
\frac{1}{\beta} \sum_{p_n} \int \frac{d^2p}{(2 \pi)^2}\,
\left(e^2 |A_x(ip_n,{\bf p})|^2 + \sum_{\nu}^{x,y,z} |A_x^{\nu}(ip_n,{\bf
p})|^2  \right) \nonumber \\ && \times \frac{1}{\beta}
\sum_{\sigma, \omega_n} \int \frac{d^2k}{(2 \pi)^2}
[\cos (k_x a) + \cos((k_x-p_x)a)] G_\sigma(i\omega_n,{\bf k}),
\end{eqnarray}
where $G^{-1}_\sigma(i\omega_n,{\bf
k})=i\omega_n+2t_a\cos(k_xa)+2t_b\cos(k_yb)+\mu$ ($\mu$ being the
Fermi level). Again, we neglect the effect of the gap
$\Delta_0$. Neglecting corrections of order $t_b/t_a$, we have
\begin{eqnarray}
\frac{1}{\beta}
\sum_{\sigma, \omega_n} \int \frac{d^2k}{(2 \pi)^2}
[\cos (k_x a) + \cos((k_x-p_x)a)] G_\sigma(i\omega_n,{\bf k}) &=&
\frac{2}{\pi b} \int_{-\pi/a}^{\pi/a} \frac{dk_x}{2\pi} \cos(k_xa)
\Theta(2t_a\cos(k_xa)+\mu) = v_F^2N(0) ,
\end{eqnarray}
which yields
\begin{equation}
\langle S_x^{\rm dia} \rangle_{\rm MF} = N(0)v_{F}^2
\frac{1}{\beta} \sum_{p_n} \int \frac{d^2p}{(2 \pi)^2}
\left(e^2 |A_x(ip_n,{\bf p})|^2 + \sum_{\nu}^{x,y,z}
|A_x^{\nu}(ip_n,{\bf p})|^2 \right) .
\label{xdiaf}
\end{equation}
\eleq
We do not consider the term of order $q_x^2$, which is consistent with
the linearized dispersion law used in the rest of the calculation.

\subsubsection{Chiral anomaly}
\label{sssecca}

In this section, we calculate the action $S_J$ due to the Jacobian of
the chiral transformation (\ref{Uchiral}). The latter produces not
only a change of the gauge fields $A_\mu^{\rm tot}$ [Eqs.~(\ref{tgf2})
and (\ref{tgf3})], but also changes the ground state of the
system. This non-perturbative effect shows up in the Jacobian of the
chiral transformation. Chiral anomalies have been known for a long
time in the context of DW
systems. \cite{Krive85,Su86,Ishikawa88,Nagaosa96,Yak98} Our method of
calculation is similar to that of Ref.~\onlinecite{Yak98}.

The chiral transformation changes the local density of particle in the
ground state (the total particle number remaining unchanged). Since
the particle density couples to the gauge field $A_0^0$, this yields
an additional contribution, $S_J$, to the action.

Let us first calculate the density change $\delta\rho(\tau,{\bf r})$
due to an infinitesimal chiral transformation,
\begin{equation}
U_{\rm chiral}[\delta\theta(\tau,{\bf r})]
=e^{\frac{i}{2}\tau_3\delta\theta(\tau,{\bf r})},
\end{equation}
which changes the phase of the order parameter from $\theta(\tau,{\bf
r})$ to $\theta(\tau,{\bf r})-\delta\theta(\tau,{\bf r})$:
\bleq
\begin{equation}
\delta \rho(\tau,{\bf r}) = \lim_{\delta{\bf r},\delta \tau \to 0}
\left \langle\phi^{\dagger}(\tau + \delta \tau, {\bf r}+\delta {\bf r})
\left( U^{\dagger}_{\rm chiral}[\delta \theta(\tau + \delta \tau, {\bf r}
+\delta {\bf r})]
U_{\rm chiral}[\delta \theta(\tau,{\bf r})] - 1 \right)
\phi(\tau, {\bf r}) \right \rangle,
\end{equation}
where we use the point spitting regularization scheme. A
regularization is necessary to properly calculate the particle
density.\cite{Yak98} To lowest order in $\delta\theta$, we obtain
\begin{eqnarray}
\delta \rho(\tau,{\bf r}) &=& -\frac{i}{2} \lim_{\delta{\bf r},\delta \tau
\to 0} [\delta\theta(\tau+\delta\tau,{\bf r}+\delta{\bf r})-\delta
\theta(\tau,{\bf r})] \langle \phi^\dagger(\tau+\delta\tau,{\bf r}+\delta{\bf
r}) \tau_3 \phi(\tau,{\bf r}) \rangle \nonumber \\
&=& -\frac{i}{2} \lim_{\delta{\bf r},\delta \tau
\to 0} [\delta\theta(\tau+\delta\tau,{\bf r}+\delta{\bf r})-\delta
\theta(\tau,{\bf r})] \sum_{\alpha,\sigma} \alpha G_{\alpha\sigma}
(-\delta\tau,-\delta{\bf r}),
\end{eqnarray}
where $G_{\alpha\sigma}$ is the mean-field propagator defined in
Sec.~\ref{mft}.\cite{note4} Introducing the Fourier transforms
$\delta\theta(ip_n,{\bf p})$ and $G_{\alpha\sigma}(i\omega_n,{\bf
k})$, $\delta\rho$ is rewritten as
\begin{equation}
\delta\rho(\tau,{\bf r}) = -\frac{i}{2\beta^2} \sum_{p_n,\omega_n} \int
\frac{d^2p}{(2\pi)^2}  \int \frac{d^2k}{(2\pi)^2}\, \delta\theta(ip_n,{\bf
p}) e^{i{\bf p}\cdot {\bf r} -ip_n\tau} \sum_{\alpha,\sigma} \alpha
[G_{\alpha\sigma}(i\omega_n+ip_n,{\bf k}+{\bf p})
-G_{\alpha\sigma}(i\omega_n,{\bf k})],
\end{equation}
after a trivial shift of integration variables, and in the limit
$\delta\tau= \delta{\bf r}=0$. Performing the sum over $\omega_n$, we
obtain
\begin{equation}
\delta\rho(\tau,{\bf r}) = -\frac{i}{\beta}\sum_{p_n} \int
\frac{d^2p}{(2\pi)^2}  \int \frac{d^2k}{(2\pi)^2}\, \delta\theta(ip_n,{\bf
p}) e^{i{\bf p}\cdot {\bf r} -ip_n\tau} \sum_\alpha \alpha
[n_\alpha({\bf k}+{\bf p})-n_\alpha({\bf k})] ,
\end{equation}
where $n_\alpha({\bf k})=\Theta (k_F-\alpha k_x+(2t_b/v_F)\cos(k_yb))$
is the particle occupation number. [Note that $n_\alpha({\bf k})$ has
the same value in the metallic and SDW phases at $T=0$.] Using
\begin{equation}
n_\alpha({\bf k}+{\bf p})-n_\alpha({\bf k}) = -\Bigl (\alpha
p_x+\frac{2t_b}{v_F}p_y \sin(k_yb) \Bigr ) \delta\Bigl(k_F-\alpha k_x+
\frac{2t_b}{v_F} \cos(k_yb)\Bigr)
\end{equation}
\eleq
we finally obtain, to lowest order in ${\bf p}$,
\begin{eqnarray}
\delta\rho (\tau,{\bf r}) &=& \frac{i}{\pi b\beta} \sum_{p_n} \int
\frac{d^2p}{(2\pi)^2} p_x \delta\theta(ip_n,{\bf
p}) e^{i{\bf p}\cdot {\bf r} -ip_n\tau} \nonumber \\
&=& \frac{1}{\pi b} \partial_x \delta\theta (\tau,{\bf r}) .
\label{Drho}
\end{eqnarray}

By integrating out the Fermions, we have shown that the ground-state
density of Fermions couples to the gauge field $A_0^0$
[Eq.~(\ref{lt1})]. Since an infinitesimal chiral transformation
changes the density by $\delta\rho$, it also produces the contribution
to the effective action\cite{note5}
\begin{eqnarray}
\delta S_J &=& - \int_0^\beta d^2r\, d\tau\, \delta\rho A_0^0 \nonumber \\
&=& - \frac{1}{\pi b} \int_0^\beta d^2r\, d\tau\, \partial_x \delta\theta
\Bigl (eA_0+i\delta\rho^{\rm HS}-\frac{v_F}{2}\partial_x \theta \Bigr
) . \nonumber\\
\label{SJ}
\end{eqnarray}
Taking a variational integral of Eq.~(\ref{SJ}), we obtain the chiral
anomaly contribution to the effective action
\begin{eqnarray}
S_J &=& -v_FN(0)\int_0^\beta d^2r\, d\tau\, \Bigl [ \partial_x \theta
(eA_0 + i\delta\rho^{\rm HS}) \nonumber\\
&& - \frac{v_F}{4}
(\partial_x\theta)^2 \Bigr ],
\label{chac2}
\end{eqnarray}
where we have used $1/\pi b=v_FN(0)$.

In deriving the action $S_J$, we have implicitly assumed that the
chiral transformation only produces a change in the density of the
ground state. Following analogous steps leading to Eq.~(\ref{Drho}),
it can be shown that the chiral transformation does not change the
spin density and the spin or charge currents, so that the change in
the density is the only non-perturbative effect of the chiral
transformation.

\subsubsection{ $\langle S''^2\rangle_{\rm MF}$}
\label{sssecsot}

In this section, we calculate $\langle S''^2\rangle_{\rm MF}$. Using
(\ref{eac0}), we deduce
\bleq
\begin{eqnarray}
\langle S''^2\rangle_{\rm MF}
&=&
\frac{1}{\beta} \sum_{p_n} \int \frac{d^2p}{(2\pi)^2}
\,\Big[  \sum_{\alpha\,\beta}^{0,x,y,z} \sum_{\mu\,\nu }^{0,x,y}
A_{\mu}^{\alpha}(ip_n,{\bf p})
 \Pi_{j_{\mu}^{\alpha}j_{\nu}^{\beta}}(ip_n,{\bf p})
A_{\nu}^{\beta}(-ip_n,-{\bf p}) \nonumber\\
&& +  \sum_{\alpha}^{0,x,y,z} \sum_{\mu}^{0,x,y}
A_{\mu}^{\alpha}(ip_n,{\bf p})
\Pi_{j_{\mu}^{\alpha} g}(ip_n,{\bf p})
ip_y \theta(-ip_n,-{\bf p}) + \frac{p_y^2}{4} |\theta(ip_n,{\bf p})|^2
\Pi_{g g}(ip_n,{\bf p})  ,
\label{eac3}
\end{eqnarray}
\eleq
where $\Pi_{j_{\mu}^{\alpha}j_{\nu}^{\beta}}(ip_n,{\bf p})$,
$\Pi_{j_{\mu}^{\alpha} g }(ip_n,{\bf p})$ and $\Pi_{g g}(ip_n,{\bf
p})$ are the current-current correlators defined by
\begin{eqnarray}
\Pi_{j_{\mu}^{\alpha} j_{\nu}^{\beta}}(ip_n,{\bf p}) &=&
\left \langle j_{\mu}^{\alpha}(ip_n,{\bf p})
j_{\nu}^{\beta}(-ip_n,-{\bf p}) \right \rangle_{\rm MF},\nonumber\\
\Pi_{j_{\mu}^{\alpha} g}(ip_n,{\bf p}) &=&
\left \langle j_{\mu}^{\alpha}(ip_n,{\bf p})
g(-ip_n,-{\bf p}) \right \rangle_{\rm MF} ,\nonumber \\
\Pi_{g g}(ip_n,{\bf p}) &=&
\left \langle g(ip_n,{\bf p}) g(-ip_n,-{\bf p})
\right \rangle_{\rm MF},
\end{eqnarray}
and the currents $j_{\mu}^{\alpha}(ip_n,{\bf p})$ and $g(ip_n,{\bf
p})$ are the Fourier transforms of the charge, spin and chiral
currents defined in Eq.~(\ref{scc}).

In the long-wavelength limit, we need to compute the effective action
to second order in a gradient expansion. For symmetry reasons, the
correlators $\Pi_{j_{\mu}^{x,y} j_{\mu'}^{0,z}}$, $\Pi_{j_{\mu}^{x,y}
g}$, $\Pi_{j_{\mu}^x j_{\mu'}^{y}}$, and $\Pi_{j_{\mu}^z g}$ vanish,
and $\Pi_{j_{\mu}^x j_{\mu'}^{x}} = \Pi_{j_{\mu}^y j_{\mu'}^{y}}$. We
also find that none of the spin currents $j_\mu^{x,y,z}$ couple to the
charge currents $j^0_\mu$ or $g$, which simply states that the sliding
and spin-wave modes decouple in the long-wavelength limit. Since the
SU(2) gauge fields are first order in gradient ($A_\mu^\nu \propto
\partial_\mu n^\nu$), we need $\Pi_{j_{\mu}^\nu j_{\mu'}^{\nu'}}$ to
zeroth order. This also holds for the correlator $\Pi_{gg}$, since it
couples to $(\partial_y\theta)^2$. Based on similar reasoning, we must
obtain $\Pi_{gj^0_\mu}$ to first order and $\Pi_{j^0_\mu j^0_{\mu'}}$
to second order. As shown in the Appendix, we thus obtain
\begin{eqnarray}
\Pi_{j_{0}^0 j_{0}^{0}} (ip_n,{\bf p}) &=&  N(0) \frac{v_F^2 p_x^2 +
v_{\perp}^2 p_y^2}{3\Delta_0^2}, \nonumber\\
\Pi_{j_{x}^0 j_{x}^{0}}(ip_n,{\bf p})  &=& 2N(0)v_F^2
\left(1-\frac{p_n^2}{6\Delta_0^2} \right), \nonumber\\
\Pi_{j_{y}^0 j_{y}^{0}}(ip_n,{\bf p})  &=& 2N(0)v_{\perp}^2
\left(1-\frac{p_y^2 b^2}{4}-\frac{p_n^2}{6\Delta_0^2} \right), \nonumber\\
\Pi_{j_{0}^0 j_{x}^{0}}(ip_n,{\bf p})  &=& N(0)v_F^2
\frac{i p_n p_x}{3\Delta_0^2}, \nonumber\\
\Pi_{j_{0}^0 j_{y}^{0}}(ip_n,{\bf p})  &=& N(0)v_{\perp}^2
\frac{i p_n p_y}{3\Delta_0^2}, \nonumber\\
\Pi_{j_{0}^x j_{0}^{x}}(ip_n,{\bf p})  &=&\Pi_{j_{0}^y j_{0}^{y}}(ip_n,{\bf
p}) = 2N(0),\nonumber\\
\Pi_{j_{x}^z j_{x}^{z}}(ip_n,{\bf p})  &=& 2N(0)v_F^2 ,\nonumber\\
\Pi_{j_{y}^z j_{y}^{z}}(ip_n,{\bf p})  &=& 2N(0)v_{\perp}^2.
\label{nvc}
\end{eqnarray}

Separating charge and spin contributions, we have $-\langle
S''^2\rangle_{\rm MF}/2= S_{2\,{\rm spin}}^{\rm eff} + S_{2\,{\rm
phason}}^{\rm eff}$, where
\bleq
\begin{eqnarray}
S_{2\,{\rm spin}}^{\rm eff} &=& -N(0)\frac{1}{\beta} \sum_{p_n} \int
\frac{d^2p}{(2\pi)^2} \,\big[ |A_0^x(ip_n,{\bf p})|^2+
|A_0^y(ip_n,{\bf p})|^2 + v_F^2  |A_x^z(ip_n,{\bf p})|^2 + v_{\perp}^2
|A_y^z(ip_n,{\bf p})|^2 \big] ,\nonumber\\
S_{2\,{\rm phason}}^{\rm eff} &=&  -N(0)\frac{1}{\beta} \sum_{p_n} \int
\frac{d^2p}{(2\pi)^2}\,{\Bigg\{}|eA_0(ip_n,{\bf p}) + i\delta
\rho^{\rm HS}(ip_n,{\bf p})|^2 \left(\frac{v_F^2 p_x^2
+ v_{\perp}^2 p_y^2}{6 \Delta_0^2}\right)
+ e^2|A_x(ip_n,{\bf p})|^2 v_F^2\left( 1-\frac{p_n^2}{6 \Delta_0^2}
\right)  \nonumber\\ &&
+ e^2|A_y(ip_n,{\bf p})|^2 v_{\perp}^2 \left( 1- \frac{p_y^2 b^2}{4}
- \frac{p_n^2}{6 \Delta_0^2} \right)
- \frac{1}{4}|\theta(ip_n,{\bf p})|^2 p_n^2
-\theta(ip_n,{\bf p}) e v_F p_n A_x(-ip_n,-{\bf p}) \nonumber \\
&& + \frac{ip_n}{3 \Delta_0^2}
\left[eA_0(ip_n,{\bf p}) + i\delta \rho^{\rm HS}(ip_n,{\bf
p})\right] \left[v_F^2 p_x  eA_x(-ip_n,-{\bf p}) + v_{\perp}^2 p_y
eA_y(-ip_n,-{\bf p})\right] {\Bigg \}}.
\label{eac4}
\end{eqnarray}
\eleq

This completes our derivation of the effective action. In the next
sections, we shall use this effective action to obtain the spin-wave
and the phason modes.

\section{Spin-wave mode}
\label{secswm}

\subsection{NL$\sigma$M}
\label{ssecnlsm}

The contribution to the spin part $S^{\rm eff}_{\rm spin}$ of the
effective action comes from the terms involving the gauge fields
$A_{\mu}^{x,y,z}$. From Eqs.~(\ref{ydiaf}), (\ref{xdiaf}) and
(\ref{eac4}), we find
\begin{eqnarray}
S^{\rm eff}_{\rm spin} &=& -N(0)\sum_{\nu}^{x,y}\frac{1}{\beta}
\sum_{p_n} \int \frac{d^2p}{(2\pi)^2}\,
\big[ |A_0^{\nu}(ip_n,{\bf p})|^2\nonumber\\
&& - v_F^2 |A_x^{\nu}(ip_n,{\bf p})|^2  - v_{\perp}^2 |A_y^{\nu}(ip_n,{\bf
p})|^2  \big].
\label{seac1}
\end{eqnarray}
The contribution from the diamagnetic terms [Eqs.~(\ref{ydiaf}) and
(\ref{xdiaf})] cancels the terms proportional to $A_{\mu}^z$ in
Eq.~(\ref{eac4}), so that we are finally left with an effective action
which depends only on $A_{\mu}^x$ and $A_{\mu}^y$. This cancelation is
expected since our effective action has to be gauge invariant. As
mentioned earlier, for the SU(2) fields $C_{\mu}$ a gauge
transformation corresponds to an arbitrary rotation about the $z$
axis, and does not change the state of the system. Such a rotation
changes $A_{\mu}^z \to A_{\mu}^z + 1/2 \, \partial_{\mu}\Lambda$,
where $\Lambda$ is the rotation angle. \cite{Schakel98} Thus to be
gauge-invariant, the effective action cannot depend on the $A_{\mu}^z$
field. The diamagnetic terms in the effective action are absolutely
crucial for this cancelation and hence for obtaining a gauge-invariant
effective action for the spin-wave mode.

Using the identities \cite{Schakel98}
\begin{eqnarray}
{\rm tr}(C_{(x,y)}^2) - {\rm tr}(\sigma_z C_{(x,y)})^2
&=& {A^x_{(x,y)}}^2+ {A^y_{(x,y)}}^2 \nonumber\\ &=&
\frac{(\partial_{(x,y)} {\bf n})^2}{4} ,\nonumber\\
{\rm tr}(C_{0}^2) - {\rm tr}(\sigma_z C_{0})^2
&=&  {A^x_0}^2+ {A^y_0}^2 =
-\frac{(\partial_{\tau} {\bf n})^2}{4}, \nonumber\\
\end{eqnarray}
we express the effective action [Eq.~(\ref{seac1})] in terms of
gradients of the ${\bf n}$ field:
\begin{equation}
S^{\rm eff}_{\rm spin} = \frac{N(0)}{4}\int_0^{\beta} d^2 r d\tau
\left[(\partial_{\tau} {\bf n})^2 + v_F^2 (\partial_{x} {\bf n})^2
+v_{\perp}^2 (\partial_{y} {\bf n})^2 \right]. 
\label{seac2}
\end{equation}

The effective action for the spins therefore turns out to be an
anisotropic NL$\sigma$M, as expected. Since $S^{\rm eff}_{\rm spin}$
has been obtained within a gradient expansion, valid for
$|p_n|,v_F|p_x|,v_\perp|p_y|\ll \Delta_0$, the NL$\sigma$M describes
the behavior of the model only in the low-energy long-wavelength
limit. Thus, Eq.~(\ref{seac2}) should be supplemented with cut-offs
$\Lambda_0\sim \Delta_0$ in energy space and $\Lambda_x\sim
\Delta_0/v_F$ and $\Lambda_y\sim \Delta_0/v_\perp$ in momentum space.

The action (\ref{seac2}) can be expressed in a more conventional
form\cite{Auerbach} as
\begin{equation}
S^{\rm eff}_{\rm spin} = \frac{1}{2}\int_0^{\beta} d^2 rd\tau
\left[\chi (\partial_{\tau} {\bf n})^2 + \rho_{sx}(\partial_{x} {\bf n})^2
+\rho_{sy}(\partial_{y} {\bf n})^2 \right],
\label{seac3}
\end{equation}
where we have introduced the parameters $\chi = N(0)/2$, $\rho_{sx} =
N(0)v_F^2/2$ and $\rho_{sy} = N(0)v_{\perp}^2/2$. Note that these
values are only approximate. In order to obtain the correct low-energy
long-wavelength behavior, it would be necessary to integrate out all
high-energy short-wavelength fluctuations. The latter are expected to
renormalize the bare values of $\chi$, $\rho_{sx}$ and $\rho_{sy}$
quoted above.

To see the physical interpretation of the parameter $\chi$, we note
that $\chi = \Pi_{j_{0}^{\nu} j_{0}^{\nu}}(ip_n=0,{\bf p}=0)/4$
($\nu=x,y$).  The correlators $\Pi_{j_{0}^{x} j_{0}^{x}}$ and
$\Pi_{j_{0}^{y} j_{0}^{y}}$ can be linked to the transverse spin
susceptibilities in the mean-field state, {\it i.e.}
$\Pi_{j_{0}^{x(y)} j_{0}^{x(y)}} = 4 \langle S_{x(y)}S_{x(y)}
\rangle_{\rm MF}$, where $S_{\mu} = \phi^{\dagger}\sigma_{\mu}\phi
/2$. We see that $\chi =\langle S_{x}S_{x} \rangle_{\rm MF} = \langle
S_{y}S_{y}\rangle_{\rm MF}$ and hence corresponds to the transverse
static uniform spin-susceptibility of the system in the mean-field
ground state.

The spin stiffness coefficients $\rho_{sx}$ and $\rho_{sy}$, on the
other hand, come from the diamagnetic terms [Eqs.~(\ref{ydiaf}) and
(\ref{xdiaf})], which are themselves related to the average kinetic
energy. To see this point more clearly, let us consider the
diamagnetic term in the $y$ direction. We have
\bleq
\begin{eqnarray}
N(0) v_{\perp}^2 \sum_{\mu}^{x,y}\frac{1}{\beta} \sum_{p_n} \int
\frac{d^2p}{(2\pi)^2}  \,|A_y^{\mu}(ip_n,{\bf p})|^2
 &=& \frac{t_b b^2}{2} \sum_{\mu}^{x,y}\frac{1}{\beta} \sum_{p_n} \int
\frac{d^2p}{(2\pi)^2}  |A_y^{\mu}(ip_n,{\bf p})|^2
\frac{1}{\beta} \sum_{i\omega_n,\alpha,\sigma} \frac{d^2k}{(2\pi)^2}
2 \cos(k_yb) G_{\alpha\sigma}(i\omega_n,{\bf k}) \nonumber\\
&=& -\frac{b^2}{2} \langle K_y \rangle_{\rm MF}
\sum_{\mu}^{x,y}\frac{1}{\beta} \sum_{p_n} \int
\frac{d^2p}{(2\pi)^2}  |A_y^{\mu}(ip_n,{\bf p})|^2 ,
\label{ake}
\end{eqnarray}
where $\langle K_y \rangle_{\rm MF}$ is the average kinetic energy in
the $y$ direction per unit area in the mean-field state. But we can
also write the diamagnetic term as
\begin{eqnarray}
N(0) v_{\perp}^2 \sum_{\mu}^{x,y}\frac{1}{\beta} \sum_{p_n} \int
\frac{d^2p}{(2\pi)^2}  \,|A_y^{\mu}(ip_n,{\bf p})|^2
&=& \frac{\rho_{sy}}{2} \int_0^{\beta} d\tau d^2 r
\left(\frac{\partial {\bf n}}{\partial y} \right)^2 .
\label{spst}
\end{eqnarray}
\eleq
Comparing Eqs.~(\ref{ake}) and (\ref{spst}), we immediately see that
the spin stiffness parameter $\rho_{sy}$ is related to the average
kinetic energy in the $y$ direction. A similar result can be obtained
for $\rho_{sx}$, and we finally get
\begin{eqnarray}
\rho_{sx} &=& -\frac{a^2 \langle K_{x} \rangle_{\rm MF}}{4}, \nonumber\\
\rho_{sy} &=& -\frac{ b^2 \langle K_{y} \rangle_{\rm MF}}{4}.
\end{eqnarray}
The velocities of the spin-wave mode, $v_x=(\rho_{sx}/\chi)^{1/2}$ and
$v_y=(\rho_{sy}/\chi)^{1/2}$, are then given by
\begin{eqnarray}
v_x &=& \left(\frac{-a^2 \langle K_{x}
\rangle_{\rm MF}}{4 \chi} \right)^{1/2}=v_F, \nonumber\\
v_y &=& \left(\frac{-b^2 \langle K_{y}
\rangle_{\rm MF}}{4 \chi} \right)^{1/2}=v_\perp .
\end{eqnarray}

We find that the longitudinal velocity equals the Fermi velocity,
because we have neglected the coupling of the spin-wave mode with the
long-wavelength fluctuations. Taking into account the latter would
lead to the renormalized spin-wave velocity
$v_x=v_F(1-UN(0))^{1/2}$. \cite{Gruner94}

Finally, we note that we can carry out an appropriate rescaling of
lengths to obtain an isotropic NL$\sigma$M. Rescaling the lengths as
$x'= x/v_x$, $y'=y/v_y$ and $\tau'=\tau$, we obtain
\begin{eqnarray}
S^{\rm eff}_{\rm spin} &=& \frac{\sqrt{\rho_{sx}\rho_{sy}}}{2}
\int_{\Lambda'} d^2 r'\, d\tau'
\sum_{\mu}^{\tau',x',y'} (\partial_{\mu} {\bf n})^2.
\label{seac5}
\end{eqnarray}
The cut-off $\Lambda'\sim \Delta_0$ (in reciprocal space) is now
isotropic.  The dimensionless coupling constant of the isotropic
NL$\sigma$M (\ref{seac5}) equals $\Delta_0/(\rho_{sx}
\rho_{sy})^{1/2}$.  It is proportional to $\Delta_0\propto
e^{-2/UN(0)}$ and hence approaches zero in the weak-coupling limit ($U
\to 0$).\cite{note3}

\subsection{Berry phase term}
\label{ssecbpt}

In general, in addition to the usual dynamical terms, the action
describing antiferromagnetic spin systems in one or two dimensions is
expected to have a topological Berry phase term.\cite{Auerbach} Such a
term was derived for antiferromagnetic systems described by the 2D
isotropic Hubbard model by Wen ${\it et\,al.}$ \cite{Wen88} We now
derive this term for our action.

The Berry phase term comes from the term proportional to $A_0^z$ in
$\langle S''\rangle_{\rm MF}$. However, to get a non-zero result, it
is not sufficient to retain only the long-wavelength part of the $R$
matrix, since the Berry phase term precisely results from the $2k_F$
oscillating part. Instead of considering the more general structure of
the $R$ matrix, we go back to the original tight-binding dispersion
law, which is a more direct way for obtaining the Berry phase
term. The term proportional to $A_0^z$ in $\langle S''\rangle_{\rm
MF}$ then reads
\begin{eqnarray}
S_{\rm Berry} &=& - \int d^2r \,d\tau \,A_0^z \,\langle \phi^{\dagger}
\sigma_z \phi \rangle_{\rm MF},
\end{eqnarray}
where $\langle \phi^{\dagger}\sigma_z \phi \rangle_{\rm MF}$ is the
spin density in the mean-field state. Noting that $\langle
\phi^{\dagger}\sigma_z \phi \rangle_{\rm MF}$ becomes $\langle
\phi^{\dagger}(\tau_0+\tau_+ + \tau_-) \sigma_z \phi \rangle_{\rm MF}$
in the continuous formulation, we obtain, using the mean-field
equations [Eq.~(\ref{mfe})],
\begin{eqnarray}
S_{\rm Berry} &=& - \int d^2r \,d\tau \,A_0^z
\langle \phi^{\dagger}(\tau_0 +\tau_+ +
\tau_-) \sigma_z \phi \rangle_{\rm MF} \nonumber\\
&=& -\frac{4\Delta_0}{U} \int d^2r \,d\tau \,A_0^z
\cos({\bf Q}\cdot {\bf r}).
\end{eqnarray}
In general, it is not possible to express the field $A_0^z$ in terms
of the ${\bf n}$ vector, because the former is a gauge-dependent
quantity. Nevertheless, it is still possible to express the variation
of the $A_0^z$ field in terms of ${\bf n}$ as \cite{Wen88} $ A_0^z = -
i/2 \int_{0}^1 d \eta {\bf n}(\eta) \cdot (\partial_{\eta}{\bf
n}(\eta) \times \partial_{\tau}{\bf n}(\eta))$, where $\eta$ is an
external parameter varying continuously from 0 to 1, ${\bf n}(1)$ is
the physical ${\bf n}$ field and ${\bf n}(0) = 0$. The action can then
be written as
\begin{eqnarray}
S_{\rm Berry} &=& i\frac{2\Delta_0}{U}\int d^2r \,d\tau\,
\cos({\bf Q}\cdot {\bf r}) \nonumber\\
&& \times \int_{0}^1  d\eta \,
{\bf n}(\eta) \cdot \left[\partial_{\eta}{\bf n}(\eta)
\times \partial_{\tau}{\bf n}(\eta) \right] .
\label{bt1}
\end{eqnarray}
Another equivalent way of writing the Berry term in the effective
action is to express the gauge field $A_0^z$ in terms of the polar and
azimuthal angles $\alpha$ and $\phi$ of the ${\bf n}$ vector. Using
$A_0^z = -i/2 \,(1-\cos(\alpha(\tau,{\bf r})) \partial_{\tau}
\phi(\tau,{\bf r})$, the Berry term in the action then assumes the
more familiar form \cite{Auerbach}
\begin{eqnarray}
S_{\rm Berry} &=& i\frac{2\Delta_0}{U}\int d^2r \,d\tau
\cos({\bf Q}\cdot {\bf r}) \nonumber\\
&& \times [1-\cos(\alpha(\tau,{\bf r}))]
\partial_{\tau} \phi(\tau,{\bf r}) .
\label{bt2}
\end{eqnarray}

For 2D and 3D antiferromagnets, it is generally believed that the
presence of the Berry phase term plays no role.\cite{Auerbach} The
dynamics of the spin-wave mode is therefore determined by the
NL$\sigma$M [Eq.~(\ref{seac3})] derived in the previous section.

\section{Phason mode}
\label{secpm}

The contribution to the phason effective action comes from the
diamagnetic terms [Eqs.~(\ref{ydiaf}) and (\ref{xdiaf})], the chiral
term $S_J$ [Eq.~(\ref{chac2})], the interaction term $S_I$
[Eq.~(\ref{it})] and $S_{2\,{\rm phason}}$ [Eq.~(\ref{eac4})]:
\bleq
\begin{eqnarray}
S_{\rm phason}^{\rm eff} &=&  \frac{1}{\beta} \sum_{p_n} \int
\frac{d^2p}{(2\pi)^2}\,{\Big \{} \sum_{\mu \nu}^{0,x,y}
\frac{1}{2}
\left[(eA_{\mu}(ip_n,{\bf p}) + i \delta \rho^{\rm HS}(ip_n,{\bf p})
\delta_{\mu 0})
P_{\mu \nu}(ip_n,{\bf p}) (eA_{\nu}(-ip_n,-{\bf p}) + i \delta \rho^{\rm
HS}(-ip_n,-{\bf p}) \delta_{\nu 0}) \right] \nonumber\\
&& + \sum_{\mu}^{0,x,y}
\left[(eA_{\mu}(ip_n,{\bf p})+i\delta\rho^{\rm HS}
(ip_n,{\bf p})\delta_{\mu 0})
F_{\mu}(ip_n,{\bf p}) \theta(-ip_n,-{\bf p}) \right]
+ \frac{1}{2}|\theta(ip_n,{\bf p})|^2 D(ip_n,{\bf p}) \nonumber\\
&& +i \delta \rho(ip_n,{\bf p}) \delta \rho^{\rm HS} (-ip_n,-{\bf p})
+\frac{U}{4} |\delta \rho(ip_n,{\bf p})|^2
{\Big \}},
\label{phac}
\end{eqnarray}
where the polarization tensor $P_{\mu \nu}$ and the coupling
coefficients $F_{\mu}$ and $D$, are given by
\begin{eqnarray}
P_{\mu \nu}(ip_n,{\bf p}) &=& \left( \begin{array}{ccc}
-\frac{N(0)}{3\Delta_0^2}(v_F^2 p_x^2 + v_{\perp}^2 p_y^2)&
-\frac{N(0)v_F^2}{3\Delta_0^2} ip_n p_x & -\frac{N(0)v_{\perp}^2}
{3\Delta_0^2} ip_n p_y \\
-\frac{N(0)v_F^2}{3\Delta_0^2} ip_n p_x &
\frac{N(0)v_F^2}{3\Delta_0^2} p_n^2 & 0 \\
-\frac{N(0)v_{\perp}^2}{3\Delta_0^2} ip_n p_y & 0 &
\frac{N(0)v_{\perp}^2}{3\Delta_0^2} p_n^2 \\
\end{array} \right) ,\nonumber\\
F_{\mu}(ip_n,{\bf p}) &=& \left(iN(0)v_Fp_x ,-N(0)v_Fp_n, 0
\right) ,\nonumber\\
D(ip_n,{\bf p}) &=& \frac{N(0)}{2}(p_n^2 + v_F^2 p_x^2 + v_{\perp}^2
p_y^2).
\label{coeff}
\end{eqnarray}
\eleq
The contribution from the diamagnetic terms in Eqs.~(\ref{ydiaf}) and
(\ref{xdiaf}) exactly cancels the gauge-noninvariant terms in
$S_{2\,{\rm phason}}$. In fact, it can be explicitly checked that the
polarization tensor $P_{\mu \nu}$ is transverse. Also, we note that
the phase field $\theta$ couples to the (gauge-invariant) electric
field $E_x=-p_n A_{x} + ip_x A_{0}$. This, together with the
transverse polarization tensor $P_{\mu \nu}$, ensures that if we
integrate out the fields $\delta \rho$, $\delta\rho^{\rm HS}$ and
$\theta$ to obtain an effective action $S^{\rm eff}[A_{\mu}]$, it will
be gauge-invariant. Since the coupling term between $\theta$ and $A_0$
comes from $S_J$, it turns out that the contribution from the Jacobian
of the chiral anomaly is crucial for obtaining a gauge-invariant
effective action.

\subsection{Equations of motion}
\label{sseceom}

We consider the electromagnetic field as being the external field,
{\it i.e.} we neglect the phason-polariton mode.\cite{Littlewood87}
Varying the effective action with respect to the fields $\theta$,
$\rho$ and $\rho^{\rm HS}$, we obtain the three coupled equations
\begin{eqnarray}
D(ip_n,{\bf p})\theta(ip_n,{\bf p}) + \sum_{\mu}^{0,x,y}
F_{\mu}(ip_n,{\bf p}) (eA_{\mu}(ip_n,{\bf p}) && \nonumber \\
+ i \delta \rho^{\rm HS}(ip_n,{\bf p}) \delta_{\mu 0}) &=& 0, \nonumber \\
 \label{eom1} \\
F_{0}(-ip_n,-{\bf p}) \theta(ip_n,{\bf p}) + \delta \rho(ip_n,{\bf p})
&& \nonumber \\
+\sum_{\mu}^{0,x,y} P_{0 \mu}(ip_n,{\bf p}) (eA_{\mu}(ip_n,{\bf p}) + i \delta
\rho^{\rm HS}(ip_n,{\bf p}) \delta_{\mu 0} ) &=& 0, \nonumber \\
\label{eom2} \\
\frac{U}{2} \delta \rho (ip_n,{\bf p}) + i \delta \rho^{\rm
HS}(ip_n,{\bf p}) = 0. &&  \nonumber \\
\label{eom3}
\end{eqnarray}

Using Eq.~(\ref{eom3}), we can eliminate the HS field $\delta\rho^{\rm
HS}$. This gives two coupled equations for the physical fields $\delta
\rho$ and $\theta$. Substituting the expressions for $D$, $P_{\mu
\nu}$ and $F_{\mu}$ from Eq.~(\ref{coeff}), and retaining terms up to
second order in frequency and momenta, we get
\begin{eqnarray}
(p_n^2 +v_F^2 p_x^2 + v_{\perp}^2 p_y^2 ) \theta(ip_n,{\bf p}) &&
\nonumber \\ - iv_F p_x U\delta \rho(ip_n,{\bf p})
+ 2 e v_F E_x(ip_n,{\bf p}) &=& 0, \nonumber\\
v_F N(0)p_x \theta(ip_n,{\bf p}) + i\delta \rho(ip_n,{\bf p}) &=& 0.
\label{eom4}
\end{eqnarray}
Thus, we obtain the well-known relation between charge and phase
fluctuations:
 \begin{equation}
\delta \rho(\tau,{\bf r}) = v_F N(0) \partial_x \theta(\tau,{\bf r})
= \frac{1}{\pi b} \partial_x \theta(\tau,{\bf r}) .
\label{eom4a}
\end{equation}
From (\ref{eom4}), we finally obtain the (decoupled) equations of
motion
\begin{eqnarray}
(\partial_t^2 -v_{\rm ph}^2 \partial_x^2 - v_{\perp}^2 \partial_y^2 )
\theta(t,{\bf r}) &=& -2ev_F E_x(t,{\bf r}) , \nonumber \\
(\partial_t^2 -v_{\rm ph}^2 \partial_x^2 - v_{\perp}^2 \partial_y^2 )
\delta \rho(t,{\bf r}) &=& -\frac{2ev_F}{\pi b} \partial_x E_x(t,{\bf r}),
\end{eqnarray}
where we have Wick-rotated back to real time $t$, and $v_{\rm ph} =
v_F(1 + UN(0))^{1/2}$ is the renormalized phason velocity. This
renormalization of the longitudinal phason velocity has been obtained
previously from a more conventional approach based on Green function
calculations.\cite{Gruner94}

\subsection{Effective action $S[\theta, A_{\mu}]$}
\label{sseceata}

In this section, we derive the effective action $S^{\rm
eff}[\theta,A_{\mu}]$ for the phase mode.

If we try to integrate out the HS fields $\delta\rho$ and
$\delta\rho^{\rm HS}$ in $S^{\rm eff}_{\rm
phason}[\delta\rho,\delta\rho^{\rm HS}, \theta,A_{\mu}]$, we
immediately face the problem of inverting the coefficient $P_{0 0}$ of
the quadratic term in $\delta \rho^{\rm HS}$, since $P_{0 0}^{-1}$ is
singular at $p_n,{\bf p}= 0$. Nevertheless, we note from the results
of the last section, that this term does not contribute to the
equations of motion up to quadratic order in external momenta and
frequency. We can therefore ignore this term while integrating out the
field $\delta \rho^{\rm HS}$. As we shall see, the effective action so
obtained, although not exact, reproduces the correct equations of
motion for the phase field $\theta$.

Integration of the $\delta \rho^{\rm HS}$ fields within the
above-stated approximation then gives a factor of $\prod_{p_n,{\bf p}}
\delta( \delta\rho - i p_x \theta /\pi b)$ and we are left with the
partition function
\begin{equation}
Z = \int D\delta\rho D \theta D A_{\mu}
\delta( \delta\rho - \frac{i }{\pi b} p_x \theta )
e^{-S^{\rm eff}[\delta \rho , A_{\mu}, \theta]} ,
\end{equation}
with the action
\bleq
\begin{eqnarray}
S^{\rm eff}[\delta \rho , A_{\mu},\theta] &=&
\frac{1}{\beta} \sum_{p_n} \int
\frac{d^2p}{(2\pi)^2}\,{\Big \{} \sum_{\mu \nu}^{0,x,y}
\frac{1}{2}
\left[eA_{\mu}(ip_n,{\bf p})
P_{\mu \nu}(ip_n,{\bf p}) eA_{\nu}(-ip_n,-{\bf p}) \right] \nonumber\\
&& + \sum_{\mu}^{0,x,y}
\left[eA_{\mu} (ip_n,{\bf p})
F_{\mu}(ip_n,{\bf p}) \theta(-ip_n,-{\bf p}) \right]
+ \frac{1}{2} |\theta(ip_n,{\bf p})|^2 D(ip_n,{\bf p})
+\frac{U}{4} |\delta \rho(ip_n,{\bf p})|^2
{\Big \}}.
\label{iac}
\end{eqnarray}
Integrating out $\delta \rho$, one then gets the effective action
\begin{eqnarray}
S^{\rm eff}[\theta, A_{\mu}] &=&
\frac{1}{\beta} \sum_{p_n} \int
\frac{d^2p}{(2\pi)^2}\,{\Big \{} \sum_{\mu \nu}^{0,x,y}
\frac{1}{2}
\left[eA_{\mu}(ip_n,{\bf p})
P_{\mu \nu}(ip_n,{\bf p}) eA_{\nu}(-ip_n,-{\bf p}) \right] \nonumber\\
&& + eN(0)v_FE_x(ip_n,{\bf p}) \theta(-ip_n,-{\bf p})
+ \frac{1}{2}|\theta(ip_n,{\bf p})|^2 D'(ip_n,{\bf p}) {\Big \}} ,
\label{eact}
\end{eqnarray}
\eleq
where $D'(ip_n,{\bf p}) = N(0)(p_n^2 + v_{\rm ph}^2 p_x^2 +
v_{\perp}^2 p_y^2)/2$ is the effective inverse phason propagator.
This clearly demonstrates the modification of the phason velocity due
to the interaction between the phase field $\theta$ and the density
fluctuation $\delta \rho$.  It is easy to check, by varying
Eq.~(\ref{eact}) with respect to $\theta$, that this effective action
reproduces the same equation of motion for $\theta$ as
Eq.~(\ref{eom4}) derived in the earlier section.

\subsection{Effective action $S[\delta \rho, A_{\mu}]$}
\label{sseceara}

To obtain the effective action $S[\delta \rho, A_{\mu}]$ for the
density fluctuations $\delta \rho$, we start from the effective action
$S^{\rm eff}[\delta \rho , \theta, A_{\mu}]$ [Eq.~(\ref{iac})], and
integrate out the phase field $\theta$. One then obtains the effective
action
\bleq
\begin{eqnarray}
S[\delta \rho, A_{\mu}] &=& \frac{1}{\beta} \sum_{p_n} \int
\frac{d^2p}{(2\pi)^2}\,{\Big \{} \sum_{\mu \nu}^{0,x,y}
\frac{1}{2} \left[eA_\mu(ip_n,{\bf p})
P_{\mu \nu}(ip_n,{\bf p}) eA_\nu(-ip_n,-{\bf p}) \right] \nonumber\\
&& - e\Bigl(A_0(ip_n,{\bf p})-\frac{p_n}{ip_x}A_x(ip_n,{\bf p})\Bigr)
\delta\rho(-ip_n,-{\bf p})
+\frac{1}{2} |\delta\rho(ip_n,{\bf p})|^2 \chi_{\rho \rho}^{-1}{\Big \}},
\label{eacr}
\end{eqnarray}
\eleq
where we have introduced the density-density correlation function
\begin{eqnarray}
\chi_{\rho \rho}(ip_n,{\bf p})
&=& \langle \delta\rho(ip_n,{\bf p}) \delta\rho(-ip_n,-{\bf p})\rangle
\nonumber \\ &=& 2 N(0) \frac{v_F^2 p_x^2}{p_n^2 + v_{\rm
ph}^2 p_x^2 + v_{\perp}^2 p_y^2}.
\label{ddcf}
\end{eqnarray}
It is easy to see from the expression of the density-density
correlation function that the phason does not induce any transverse
density fluctuations.

The currents in the longitudinal and transverse directions are given
by
\begin{eqnarray}
j_x(ip_n,{\bf p}) &=& - \frac{\delta S}{\delta A_x(-ip_n,-{\bf p})}
\Biggr | _{A_{\mu} = 0} = -\frac{e p_n}{ip_x} \delta \rho
(ip_n,{\bf p}) ,\nonumber\\
j_y(ip_n,{\bf p}) &=& - \frac{\delta S}{\delta A_y(-ip_n,-{\bf
p})}\Biggr | _{A_{\mu} = 0} = 0.
\label{cu}
\end{eqnarray}
These expressions agree with the (real time) continuity equation
$\partial_t \rho + \nabla \cdot {\bf j} = 0$. Also, we see that there
is no current due to the phason mode in the $y$ direction. The
contribution to the current across the chain comes entirely from the
quasi-particle excitations.\cite{Virosztek99} From Eqs.~(\ref{eom4a})
and (\ref{cu}), we obtain (in real time)
\begin{equation}
j_x(t,{\bf r}) = -\frac{e}{\pi b} \partial_t \theta(t,{\bf r}).
\label{cu1}
\end{equation}
Note that the expressions of the currents as a function of the phase
$\theta$ [Eqs.~(\ref{eom4a}) and (\ref{cu1})] have been obtained to
lowest order in a gradient expansion. They have been recently
generalized to higher order in Ref.~\onlinecite{Rozhavsky99}.

The current-current correlators can also be obtained from the
effective action. Using Eqs.~(\ref{ddcf}) and (\ref{cu}), we find that
the current-current correlator $\chi_{j_\mu j_\nu}=\langle j_\mu
j_\nu\rangle$ can be expressed in terms of the density-density
correlation function as
\begin{eqnarray}
\chi_{j_x j_x}(ip_n,{\bf p}) &=& -\frac{e^2 p_n^2}{p_x^2}
\chi_{\rho \rho }(ip_n,{\bf p}) \nonumber\\
&=& -2N(0)e^2 \frac{v_F^2 p_n^2}{p_n^2
+ v_{\rm ph}^2 p_x^2 + v_{\perp}^2 p_y^2} ,\nonumber\\
\chi_{j_y j_y}(ip_n,{\bf p}) &=& \chi_{j_x j_y}(ip_n,{\bf p}) = 0.
\label{cuco}
\end{eqnarray}
From the expression of the current-current correlator, we then obtain
the ac conductivity $\sigma_{xx}(\omega)$ as
\begin{eqnarray}
\sigma_{xx}(\omega) &=&
-\frac{1}{p_n}\chi_{j_x j_x}(ip_n,{\bf p}=0)\Bigr|_{ip_n =\omega+i0^+}
\nonumber \\
&=& 2N(0)e^2 v_F^2 \left( i {\cal P} \left(\frac{1}{\omega} \right ) + \pi
\delta(\omega)\right),
\label{cond}
\end{eqnarray}
where $\cal P$ denotes the principal part.  The real part of the
conductivity therefore satisfies the f-sum rule
\begin{eqnarray}
\int^{\infty}_{-\infty} {\rm Re}\left[\sigma_{xx}(\omega)\right]
d\omega &=& \frac{\omega_p^2}{4},
\end{eqnarray}
where $\omega_p =(8 e^2v_F/b)^{1/2}$ is the plasma frequency. The
phason mode completely exhausts the spectral weight and we obtain the
well-known result \cite{Gruner94} that there is no contribution to the
longitudinal conductivity from the quasi-particle excitations in a
clean SDW system.

\section{Conclusion}
\label{secc}

We have derived the effective action for the low-energy collective
modes of quasi-1D SDW systems. The introduction of a fluctuating
spin-quantization axis in the functional integral allows to consider
the phason mode, the spin-wave mode, and the long-wavelength charge
fluctuations on equal footing.

We find that the spin-wave mode is governed by an isotropic
NL$\sigma$M together with a topological Berry phase term. By a
suitable length rescaling, one can obtain an isotropic NL$\sigma$M
with an effective dimensionless coupling constant
$\Delta_0/(\rho_{sx}\rho_{sy})^{1/2}$, where $\rho_{sx}$ and
$\rho_{sy}$ are the spin-stiffness coefficients. The coupling
constant, proportional to the mean-field gap $\Delta_0\propto
e^{-2/UN(0)}$, is small in the weak coupling limit $UN(0)\ll 1$.

The sliding mode is governed by an effective Lagrangian ${\cal
L}(\theta,\rho)$ which is a function of two independent fields: the
phase $\theta$ of the SDW condensate and the charge density field
$\rho$. From the coupled equations satisfied by these two variables,
we obtain the phason dynamics and its contribution to the
current-current and density-density correlation functions.

\section*{Acknowledgment}

One of the authors (KS) would like to thank Victor Yakovenko for
helpful discussions and for support during the work. Laboratoire de
Physique des Solides is associ\'e au CNRS.

\bleq

\appendix

\section{}
\label{secaa}

In this section, we sketch the calculation of the current-current
correlators. These correlators are given by
\begin{equation}
\Pi_{j_{\mu}^{\nu} j_{\mu'}^{\nu'}}(ip_n,{\bf p})
\equiv  \Pi_{j j'}(ip_n,{\bf p})
=  \left \langle j(ip_n,{\bf p})
j'(-ip_n,-{\bf p})  \right \rangle_{\rm MF}.
\label{corr1}
\end{equation}
The currents $j_{\mu}^{\nu}$ (or $j$ in the simplified notation) can be
defined using Eq.~(\ref{scc}) as
\begin{equation}
j(ip_n,{\bf p}) = \frac{1}{\sqrt \beta} \sum_{\omega_n} \int
\frac{d^2k}{(2\pi)^2} \sum_{\alpha}^{+-} \sum_{\sigma
\sigma'}^{\uparrow \downarrow}
\phi^{\dagger}_{\alpha\sigma'}(i\omega_n,{\bf k})v_{\alpha
\sigma \sigma'}(k_y, k_y+p_y) \phi_{\alpha \sigma }(i\omega_n +
ip_n,{\bf k}+{\bf p}),
\label{app2}
\end{equation}
where we have set the area of the system to unity. The functions
$v_{\alpha\sigma \sigma'}(k_y, k_y+p_y)$ are the current operators
which can be easily read off from the definition of the currents in
Eq.~(\ref{scc}). For example, for the charge current in the $x$
direction $j_{x}^{0}$, $v_{\alpha\sigma \sigma'}(k_y, k_y+p_y) = v_F
\alpha\delta_{\sigma \sigma'}$. From (\ref{app2}), we then deduce
\begin{eqnarray}
\Pi_{j j'}(ip_n,{\bf p}) &=& - \frac{1}{\beta}
\sum_{\omega_n} \int \frac{d^2k}{(2\pi)^2}
\sum_{\alpha}^{+-} \sum_{\sigma \sigma'}^{\uparrow \downarrow}
\left[v_{\alpha \sigma \sigma'}(k_y-p_y, k_y)
v^{'}_{\alpha \sigma' \sigma}(k_y, k_y-p_y) G_{\alpha
\sigma'}(i\omega_n,{\bf k}) G_{\alpha
\sigma}(i\omega_n-ip_n,{\bf k}-{\bf p})  \right. \nonumber\\
&& \left. + v_{\alpha \sigma \sigma'}(k_y-p_y, k_y)
v^{'}_{{\bar \alpha} \sigma' \sigma}(k_y +\pi/b, k_y-p_y +\pi/b)
F_{\alpha
\sigma'}(i\omega_n,{\bf k}) F_{{\bar \alpha}
\sigma}(i\omega_n-ip_n,{\bf k}-{\bf p}-\alpha {\bf Q}) \right],
\label{corr2}
\end{eqnarray}
where ${\bar \alpha} =\mp$ for $\alpha = \pm$ and $G_{\pm \sigma}$ and
$F_{\pm \sigma}$ are the mean-field propagators.  We evaluate the
frequency sum and the $k_x$ integral using $\epsilon_{\alpha}({\bf
k}-\alpha{\bf Q})=-\epsilon_{\bar\alpha}({\bf k})$ and expanding the
energy dispersion $\epsilon_{\alpha}$ in powers of external momenta:
\begin{eqnarray}
\epsilon_{\alpha}({\bf k})-\epsilon_{\alpha}({\bf k}-{\bf p})
&=& \epsilon_x +\epsilon_y, \nonumber\\
\epsilon_y &=& 2t_b b p_y  \sin(k_y b)
- t_b b^2 p_y^2 \cos(k_y b) + O(p_y^3),\nonumber\\
\epsilon_x &=& \alpha v_F p_x.
\end{eqnarray}
Using these relations and Eq.~(\ref{gfc}) one can evaluate the
frequency sum and the $k_x$ integral up to second order in $p_n$ and
$\epsilon_{\rm T}=\epsilon_x+\epsilon_y$ ($O(p_n^2,\epsilon_{\rm T}^2
,p_n \epsilon_{\rm T}$)) to obtain
\begin{eqnarray}
\frac{1}{\beta}\sum_{\omega_n} \int \frac{dk_x}{2\pi}\,
F_{\alpha
\sigma'}(i\omega_n,{\bf k}) F_{{\bar \alpha}
\sigma}(i\omega_n-ip_n,{\bf k}-{\bf p}-\alpha {\bf Q}) &=&
{\rm sgn}(\sigma){\rm sgn}(\sigma')\frac{N(0)}{2}
\left( \frac{1}{2} - \frac{p_n^2+\epsilon_{\rm
T}^2}{12\Delta_0^2} \right) ,\nonumber\\
\frac{1}{\beta}\sum_{\omega_n} \int \frac{dk_x}{2\pi}\,
G_{\alpha \sigma'}(i\omega_n,{\bf k})
G_{\alpha \sigma}(i\omega_n-ip_n,{\bf k}-{\bf p})&=&
- \frac{N(0)}{2} \left( \frac{1}{2} - \frac{p_n^2 - \epsilon_{\rm
T}^2}{12\Delta_0^2} + \frac{i p_n \epsilon_{\rm T}}{6\Delta_0^2}
\right).
\label{corr3}
\end{eqnarray}
Substituting Eq.~(\ref{corr3}) in the expression for the correlator
[Eq.~(\ref{corr2})] we get
\begin{eqnarray}
\Pi_{j j'}(ip_n,{\bf p})
&=& \frac{N(0)}{2} \int \frac{dk_y}{2\pi}
\sum_{\alpha}^{+-} \sum_{\sigma \sigma'}^{\uparrow \downarrow}
\left[v_{\alpha \sigma \sigma'}(k_y-p_y, k_y)
v^{'}_{\alpha \sigma' \sigma}(k_y, k_y-p_y) \left( \frac{1}{2} -
\frac{p_n^2 - \epsilon_{\rm T}^2}{12\Delta_0^2}
+ \frac{i p_n \epsilon_{\rm T}}{6\Delta_0^2} \right)
\right. \nonumber\\
&& \left. + v_{\alpha \sigma \sigma'}(k_y-p_y, k_y)
v^{'}_{{\bar \alpha} \sigma' \sigma}(k_y +\pi/b, k_y-p_y +\pi/b)
{\rm sgn}(\sigma){\rm sgn}(\sigma')\left( - \frac{1}{2}
+ \frac{p_n^2+\epsilon_{\rm T}^2}{12\Delta_0^2} \right) \right].
\label{corr4}
\end{eqnarray}

The dependence on $k_y$ comes from $\epsilon_y$ and
$v_{\alpha\sigma\sigma'}$. The latter is either $k_y$ independent or
depends on $k_y$ through the function $f=\sin((k_y-p_y)b) + \sin(k_y
b) = 2\sin(k_y b) -p_y b \cos(k_y b) -p_y^2 b^2 \sin(k_y b)/2 +
O(p_y^3)$. To evaluate the $k_y$ integral in (\ref{corr4}), we
therefore need only a limited set of integrals, which, up to order
$p_y^2$, are given by
\begin{eqnarray}
\langle f \rangle_{k_y}&=& \langle \epsilon_y \rangle_{k_y} =
\langle f\epsilon_y^2 \rangle_{k_y} =
\langle f^2 \epsilon_y \rangle_{k_y} = 0 ,\nonumber\\
\langle f^2 \rangle_{k_y} &=& 2 -\frac{ p_y^2 b^2}{2}, \,\,\,
\langle \epsilon_y^2 \rangle_{k_y} = v_{\perp}^2 p_y^2 , \,\,\,
\langle f \epsilon_y \rangle_{k_y} = 2t_b b p_y , \, \, \,
\langle f^2 \epsilon_y^2 \rangle_{k_y} = 3v_{\perp}^2 p_y^2,
\label{kysum}
\end{eqnarray}
where we have used the short-hand notation $\langle \cdots \rangle_{k_y}$
for $\int dk_y/(2\pi)\cdots$.

From Eqs.~(\ref{corr4}) and (\ref{kysum}), it is then a matter of
straightforward algebra to obtain the different correlators. As an
illustrative example we sketch the calculation of $\Pi_{j_y^0
j_y^0}$. Since the current operator in this case is given by
$v_{\alpha \sigma \sigma'}(k_y,k_y-p_y) = t_b b \delta_{\sigma
\sigma'} f(k_y,p_y)$, we obtain
\begin{eqnarray}
\Pi_{j_y^0 j_y^0}(ip_n,{\bf p}) &=& \frac{N(0) t_b^2 b^2}{2}
\sum_{\alpha}^{+-} \sum_{\sigma \sigma'}^{\uparrow \downarrow}
\delta_{\sigma \sigma'} \left \langle f^2 \left( 1-
\frac{p_n^2}{6\Delta_0^2} + \frac{ip_n \epsilon_{\rm T}}{6 \Delta_0^2}
\right )\right\rangle_{k_y} \nonumber\\
&=& 2 N(0) v_{\perp}^2 \left (1 -\frac{p_y^2 b^2}{4}
- \frac{p_n^2}{6 \Delta_0^2} \right),
\end{eqnarray}
where we have used Eq.~(\ref{kysum}) to evaluate the $k_y$ integrals
to obtain the last line and retained only terms up to second order in
external frequency and momenta.

The other correlators can be obtained following an exactly similar
procedure and in this way, we finally obtain Eqs.~(\ref{nvc}) for the
charge and spin current-current correlators.

\eleq

\ecols


\begin{thebibliography}{99}


\bibitem{Gruner94} For reviews on DW systems, see G. Gr\"uner, {\it
Density Waves in Solids} (Addison-Wesley, New York, 1994);
Rev. Mod. Phys. {\bf 60}, 1129 (1988); Rev. Mod. Phys. {\bf 66}, 1
(1994).

\bibitem{Frohlich54} H. Fr\"ohlich, Proc. Roy. Soc. Lond. A {\bf 223},
296 (1954).

\bibitem{Peierls55} R.E. Peierls, {\it Quantum Theory of Solids}
(Oxford University Press, New York, 1955).

\bibitem{Overhauser62} A.W. Overhauser, Phys. Rev. {\bf 128}, 1437
(1962).

\bibitem{Lee74} P.A. Lee, T.M. Rice, and P.W. Anderson, Solid State
Comm. {\bf 14}, 703 (1974).

\bibitem{Fukuyama76} H. Fukuyama, J. Phys. Soc. Japan {\bf 41}, 513
(1976).

\bibitem{Brazovskii76} S. Brazovski and I. Dzyaloshinskii, JETP {\bf
44}, 1233 (1976). Note that these authors perform a chiral
transformation without computing the Jacobian (chiral anomaly) so that
some parameters in their Lagrangian have incorrect values. [The chiral
anomaly ($\it{i.e.}$ the existence of a non trivial Jacobian of the
chiral transformation) was not known at that time.]

\bibitem{Fukuyama78} H. Fukuyama and P.A. Lee, Phys. Rev. B {\bf 17},
535 (1978).

\bibitem{Lee79} P.A. Lee and T.M. Rice, Phys. Rev. B {\bf 19}, 3970
(1979).

\bibitem{Psaltakis84} G.C. Psaltakis, Solid State Comm. {\bf 51}, 535
(1984).

\bibitem{Maki90a} K. Maki and A. Virosztek, Phys. Rev. B {\bf 41}, 557
(1990).

\bibitem{Maki90b} K. Maki and A. Virosztek, Phys. Rev. B {\bf 42}, 655
(1990).

\bibitem{Brazovskii93} S. Brazovskii, J. Physique I {\bf 3}, 2417
(1993).

\bibitem {Poilblanc87} D. Poilblanc and P. Lederer, Phys. Rev. B {\bf
37}, 9650 (1987); Phys. Rev. B {\bf 37}, 9672 (1987).

\bibitem{Zanchi} A. Bjelis and D. Zanchi, Phys. Rev. B {\bf 49}, 5968
(1994).

\bibitem{Maki87} K. Maki and A. Virosztek, Phys. Rev. B {\bf 36}, 511
(1987).

\bibitem{Takano82} K. Takano, Prog. Theor. Phys. {\bf 68}, 1 (1982),

\bibitem{Krive85} I.V. Krive and A.S. Rozhavsky, Phys. Lett. A{\bf
113}, 313 (1985).

\bibitem{Su86} Z. Su and B. Sakita, Phys. Rev. Lett. {\bf 56}, 780
(1986); Phys. Rev. B {\bf 38}, 7421 (1988); B. Sakita and Z. Su,
Prog. Theor. Phys. {\bf 86}, 238 (1986).

\bibitem{Ishikawa88} M. Ishikawa and H. Takayama,
Prog. Theor. Phys. {\bf 79}, 359 (1988).

\bibitem{Suzumura90} Y. Suzumura, J. Phys. Soc. Japan {\bf 59}, 1711
(1990).

\bibitem{Girard93} M. Girard, {\it M\'emoire de ma\^{\i}trise,
Universit\'e de Sherbrooke}, 1993 (unpublished).

\bibitem{Nagaosa96} N. Nagaosa and M. Oshikawa, J. Phys. Soc. Japan {\bf 65},
2241 (1996); A. Tanaka and M. Machida, J. Phys. Soc. Japan {\bf 67}, 748
(1998).

\bibitem{Yak98} V.M. Yakovenko and H.S. Goan, Phys. Rev. B {\bf 58},
10648 (1998).

\bibitem{Wen88} S. Wen and A. Zee, Phys. Rev. Lett. {\bf 61}, 1025
(1988).

\bibitem{Schulz90} H.J. Schulz, Phys. Rev. Lett. {\bf 65}, 2462
(1990); H.J. Schulz in {\it The Hubbard Model}, edited by D. Baeriswyl
${\it et \, al.}$, (Plenum Press, New York, 1995).

\bibitem{Ting91} Z.Y. Weng, C.S. Ting and T.K. Lee, Phys. Rev. B {\bf
43}, 3790 (1991).

\bibitem{note1} To be more specific, the interacting part of the
Hamiltonian can be written in the g-ology formulation as
\cite{Girard93} $H_I = \frac{1}{2} \sum_n \int dx
[(2g_1-g_2)O_n^\dagger(x)O_n(x)-g_2{\bf O}_n^\dagger(x) {\bf O}_n(x)
]$, where $O=\sum_\sigma \psi^\dagger_{-\sigma}\psi_{+\sigma}$ and
${\bf O}=\sum_{\sigma,\sigma'} \psi^\dagger_{-\sigma}\bbox{\sigma}
_{\sigma,\sigma'} \psi_{+\sigma'}$ are the charge and spin densities,
respectively (we use the notations introduced in
Sec.~\ref{secea}). $g_1=g_2=U$ in the Hubbard model. This form of the
Hamiltonian obviously preserves spin rotation invariance, and also
allows to recover the correct mean-field solution from a saddle point
approximation within a functional integral formalism.\cite{Girard93}
However, it does not take into account the long-wavelength charge or
spin fluctuations.

\bibitem{Virosztek94} It has been argued that in a SDW system the
thermally excited electrons do not screen the Coulomb interaction. As
a result, the longitudinal phason is completely absorbed by the
plasmon due to the Anderson-Higgs mechanism: A. Virosztek and K. Maki,
Phys. Rev. B {\bf 49}, 6074 (1994).

\bibitem {Palo99} S. De Palo, C. Castellani, C. Di Castro and
B.K. Chakraverty, Phys. Rev. B {\bf 60}, 564 (1999).

\bibitem{note2} Note that this limit is required for the behavior of
the system to be trully 2D. The condition $\Delta_0\gtrsim t_b$ would
imply a 1D behavior (with a weak Josephson-like interchain coupling),
a regime where the mean-field approach used in this paper breaks down.

\bibitem{Fujikawa79} K. Fujikawa, Phys. Rev. Lett. {\bf 42}, 1195
(1979).

\bibitem{note4} Note that by using the mean-field propagators, we
assume that $\delta\rho$ depends only on $\delta\theta$ (and not on
$\theta$).

\bibitem{note5} We could also expect a contribution $-\int_0^\beta
d^2r\, d\tau\, \rho\delta A_0^0$, where $\rho=\partial_x \theta$ [see
Eq.~(\ref{Drho})]. According to Ref.~\onlinecite{Yak98}, this term
does not contribute to the chiral anomaly.

\bibitem{Schakel98} A.M.J. Schakel, Cond-Mat/9805152, May 13 (1998).

\bibitem{Auerbach} A. Auerbach, {\it Interacting Electrons and Quantum
Magnetism} (Springer Verlag, New York, 1994).

\bibitem{note3} This point has also been noted by Schulz in the
context of the isotropic 2D Hubbard model. \cite{Schulz90}

\bibitem{Littlewood87} P.B. Littlewood, Phys. Rev. B {\bf 36}, 3108
(1987); S.N. Artemenko and W. Wonneberger, J. Phys. I France {\bf 6},
2079 (1996).

\bibitem{Virosztek99} A. Virosztek, B. Dora, and K. Maki,
Europhys. Lett. {\bf 47}, 358 (1999).

\bibitem{Rozhavsky99} A.S. Rozhavsky, Y.V. Pershin, and
I.A. Romanovsky, Europhys. Lett. {\bf 46}, 50 (1999).

\end{thebibliography}
\end{document}